\documentclass[prx,aps,twocolumn,superscriptaddress]{revtex4-2}

\usepackage[dvips]{graphicx}
\usepackage{amssymb,amsfonts,amsmath,gensymb,eucal}
\usepackage{color}
\usepackage[normalem]{ulem}
\usepackage{natbib}
\usepackage[hidelinks]{hyperref}
\usepackage[capitalize]{cleveref}
\crefformat{equation}{equation~(#2#1#3)}

\newcommand{\ind}[1]{\ensuremath_{\mathrm{#1}}}

\newcommand{\V}{\ensuremath{\,\mathrm{V}}}
\newcommand{\s}{\ensuremath{\,\mathrm{s}}}
\newcommand{\A}{\ensuremath{\,\mathrm{g_0V}}}
\newcommand{\RR}{\ensuremath{\,\mathrm{g_0}^{-1}}}
\newcommand{\CC}{\ensuremath{\,\mathrm{g_0s}}}
\newcommand{\LL}{\ensuremath{\,\mathrm{s}/\mathrm{g_0}}}
\newcommand{\Hz}{\ensuremath{\,\mathrm{Hz}}}

\newcommand{\Pe}{\ensuremath{\mathrm{Pe}}}
\newcommand{\Ei}{\ensuremath{\mathrm{Ei}}}

\newcommand{\dx}{\ensuremath{\mathrm{d}x}}
\newcommand{\dd}{\ensuremath{\mathrm{d}}}
\newcommand{\e}{\ensuremath{ \mathrm{e}}}

\newcommand{\nd}[2]{\ensuremath{\overline{#1 \cdot #2}}}

\newcommand{\affil}[2]{#1, Utrecht University, Princetonplein #2, 3584 CC Utrecht, The Netherlands}

\newcommand{\SCMB}{\affil{Soft Condensed Matter and Biophysics, Debye Institute for Nanomaterials Science}{1}}
\newcommand{\ITP}{\affil{Institute Theoretical Physics}{5}}

\graphicspath{{figures}}
\begin{document}

\title{Neuromorphic Computing with Microfluidic Memristors}
\author{Nex C.\ X.\ Stuhlm\"uller}\email{n.c.x.stuhlmuller@uu.nl} \affiliation{\SCMB{}}
\author{Ren\'e van Roij} \affiliation{\ITP{}}
\author{Marjolein Dijkstra}\email{m.dijkstra@uu.nl}  \affiliation{\SCMB{}}

\begin{abstract}
  Conical microfluidic channels filled with electrolytes  exhibit volatile memristive behavior, offering a promising platform for energy-efficient, neuromorphic computing.
  Here, we integrate these iontronic channels as additional nonlinear elements  in nonlinear Shinriki-inspired oscillators and demonstrate that they exhibit
  alternating chaotic and non-chaotic dynamics across  a broad
  frequency range. Exploiting this behavior, we construct  \texttt{XOR} and \texttt{NAND} gates by
  coupling three ``Memriki'' oscillators, and we further realize the full set of standard logic gates through combinations of  \texttt{NAND} gates. Our results establish a new paradigm for iontronic computing and open avenues for scalable, low-power logical operations in microfluidic and bio-inspired systems.
\end{abstract}

\maketitle

\section{Introduction}
The era we live in has been proposed  the \emph{Silicon age}~\cite{Chabal2001}, defined by semiconductor transistors that have revolutionized  data processing, accelerating it by several orders of magnitude. This technological advancement has enabled  groundbreaking inventions, including computers, the internet, and, more recently, artificial intelligence (AI).

However, advancements in AI are rapidly approaching  a critical  power bottleneck~\cite{Burr2016}.  A key factor driving this challenge is the high energy cost caused by Ohmic losses of frequent data transfer between physically separated processing and information storage units in conventional computing architectures, a limitation known as the von Neumann bottleneck~\cite{Rajendran2019}.
AI power consumption  is currently doubling every four to six months~\cite{tripp2024}, which would surpass global energy production by the 2030s.
This trend highlights the pressing need for novel  computing paradigms. To enhance energy efficiency and meet the increasing demands for computing power in the era of AI and big data, a fundamentally different approach is required~\cite{Baulin2025}.

Inspired by the most energy-efficient computer known -- the biological brain -- the field of neuromorphic computing emerged~\cite{Markovic2020}. The human brain, an intricate biological neural network comprising approximately 10$^{11}$ neurons and 10$^{15}$ synapses, operates with a remarkable efficiency at just $20\,\mathrm{W}$ of power.
A key  subfield of neuromorphic computing focuses on developing devices that mimic the computational functions of the brain. One such device is the  memristor, first theoretically proposed by Chua in 1971 as the fourth basic circuit element~\cite{Chua1971}.
Memristors act as memory elements, with a resistance that depends on  past voltage or current, allowing them to retain information. In 2008, Strukov {\em et al.} experimentally demonstrated a solid-state memristor, paving the way for its use as an artificial synapse in neuromorphic systems~\cite{Strukov2008}, with present-day applications in for instance crossbar arrays \cite{xia2019memristive,jeon2024purely}. However, silicon-based technologies typically rely on a single type of  information carrier: electrons (or their absence (holes)). In contrast, the biological brain utilizes  ions dissolved in water as information carriers, allowing for a diverse range of soluble information carriers, such as various species of ions and small molecules. A key distinction between solid-state electronics and ion-mediated biological systems lies in their energy efficiency, with biological systems operating orders of magnitude more efficiently, albeit at  much lower clock frequencies.

This contrast has led to  the emergence of iontronics, a subfield of neuromorphic computing focused on developing electronic circuit elements where ions dissolved in water serve as information carriers~\cite{Chun2015}.
Just as solid-state memristors have been widely explored, researchers have pursued iontronic analogs~\cite{Kai}. Memristive effects in charged conical nanopores were first observed in 2010 and 2012~\cite{feng2010impedance,wang2012transmembrane}, paving the way for the development of diverse nanofluidic memristors based on various mechanisms. These include  electric double layer polarization~\cite{wang2012transmembrane,kamsma2024brain,emmerich2024nanofluidic,Wang2024}, salinity gradients yielding negative differential resistance behavior~\cite{leong2020quasi}, electro-wetting~\cite{smirnov2011voltage,powell2011electric}, structural and conformational changes of  nanopores~\cite{tuszynski2020microtubules,zhou2024nanofluidic}, the ionic analogue~\cite{robin2023long} of electronic Coulomb blockade~\cite{Bose2015}, mixtures of and interfaces between ionic liquids and water~\cite{sheng2017transporting,zhang2019nanochannel}, and specific polyelectrolyte-ion interactions~\cite{xiong2023neuromorphic}.

The next step towards neuromorphic computing is integrating iontronic devices into electric circuits. Key questions include: Can these networks  emulate neural functions, perform Boolean logic operations, or execute simple machine learning tasks~\cite{Doremaele2024}?
While the use of iontronic elements in large-scale  networks is still  in its infancy, initial progress has been made in implementing basic logic gates such as \texttt{AND}, \texttt{OR}~\cite{sabbagh2023designing,Li2023},  \texttt{NAND} and \texttt{NOR}~\cite{zhang2024microscale}, and material implication \texttt{IMP}~\cite{emmerich2024nanofluidic} gates.
A combination of currents and pH as inputs have been used to realize a set of logic gates~\cite{Portillo2024,Ling2024}, requiring expensive pH changes, when switching inputs to the gate.
Nanofluidic carbon nanotubes have been used to realize nanofluidic transistors which can be used for logic gates~\cite{Liu2024}.
However, a complete set of logic gates based on (memristive) iontronics purely based on electronic inputs has yet to be realized.

In this work, we bridge iontronics and memristive computing by constructing logic gates using  microfluidic memristors~\cite{Kamsma2023}.
In addition to microfluidic memristors, we base our logic gates on Shinriki-inspired oscillators~\cite{Shinriki1981}.
Shinriki oscillators are simple electric circuits, first studied by Shinriki \textit{et al.}, that can show chaotic behavior depending on the exact system parameters.
As the human brain operates at the edge of chaos~\cite{Kitzbichler2009,chua2013memristor}, we designed our circuits to inherit this property.

We first provide a brief introduction to volatile microfluidic memristors as described in Ref.~\cite{Kamsma2023}, and
then explore  their integration into electric circuits that form Shinriki-inspired (non-linear) oscillators~\cite{Shinriki1981}. Finally, we demonstrate how these oscillators can be combined to realize the full set of logic gates, thereby advancing the use of iontronics for neuromorphic and Boolean logic applications.

\section{Microfluidic iontronic memristor}
\begin{figure}[htbp]
  \centering
  \includegraphics[width=\linewidth]{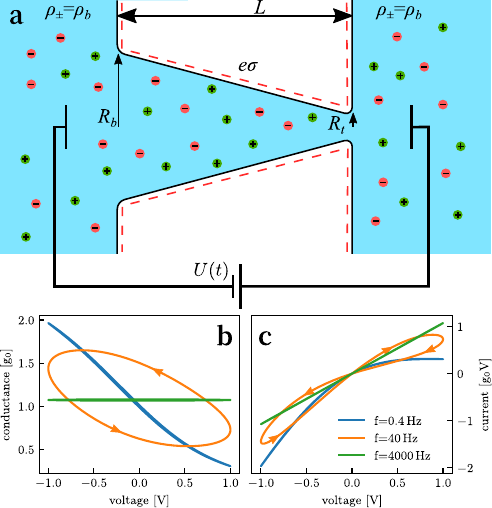}
  \caption{%
  (a) Schematic of a conical channel filled with an aqueous 1:1 electrolyte (cations in green, anions in red) with length $L$, base radius $R_b$, tip radius $R_t$, bulk ion concentrations $\rho_\pm=\rho_b$ in the reservoirs,  and surface charge $e\sigma$.
  A time-dependent voltage $U(t)$ is applied across the channel.
  (b) Limit cycles of the conductance-voltage relation for three different driving frequencies $f=0.4\Hz$ (blue), $f=40\Hz$  (orange), and $f=4000\Hz$ (green) of a sinusoidal potential with an amplitude of 1 Volt.
  (c) Limit cycles of the current-voltage relation for the same frequencies and parameters as in (b).
  }
  \label{fig:channel}
\end{figure}

In this study, we focus on volatile microfluidic memristors as presented in Ref.~\cite{Kamsma2023} and illustrated in
Fig.~\ref{fig:channel}(a). This conical channel of length $L$ has a wide opening of base radius $R_b$ at $x=0$ and a narrow tip of radius $R_t<R_b$ at $x=L$, and connects two reservoirs of an aqueous 1:1 electrolyte at a total bulk ion concentration $2\rho_b$ at room temperature. The channel can carry an ionic electric current when driven by an applied potential difference between the two reservoirs. The combination of the conical channel geometry and a negative surface charge $e\sigma$ on the channel walls results in current rectification when a  static voltage drop $U$ is applied~\cite{jubin2018dramatic}. This phenomenon was theoretically explained  in terms of a nontrivial $U$-dependent steady-state salt concentration profile $\rho_s(x;U)$ for $x\in[0,L]$ that build up within the channel~\cite{boon2022pressure}. This concentration profile is caused by the much stronger electric field near the tip compared to the base and leads to ionic depletion or accumulation depending on the polarity of the potential. Consequently, the steady-state electric conductance $g_\infty(U)$ depends non-trivially on $U$ and exhibits a hysteresis loop in the current-voltage relation when a time-dependent oscillating potential $U(t)$ is applied with a period on the order of the build-up time $\tau$ of the ionic concentration profile \cite{Kamsma2023}. The time-dependent conductance $g(t)$ of such a memristive channel was calculated from numerical solutions of the Poisson-Nernst-Planck-Stokes (partial differential) equations for the coupled transport of charge, water, and ions \cite{Kamsma2023}. In a large parameter regime $g(t)$ was also found to be accurately approximated  by the following (ordinary differential) equation of motion \footnote{this is only an approximation of the steady-state conductance, but Kamsma {\em et al.} \ found that it is reasonable within the applied voltage range~\cite{Kamsma2023,Kamsma2023a}}
\begin{gather}
\tau \dot g(t) = g_\infty(U(t)) - g(t) \label{eq:ginf} \mbox{ with}\\
g_\infty(U) = g_0\int_0^L \frac{\rho_s(x,U)}{2\rho_b L} \dx,
\end{gather}
where the functional form of the static conductance $g_\infty(U)$ is determined by the channel geometry and the electrolyte properties, that we keep fixed here however such that its numerical evaluation for different $U$ is straightforward. The parameter
$g_0$ represents the zero-field conductance, which we use as the unit of conductance below.
Throughout this work we will use the (physically realistic) set of system parameters of Ref.~\cite{Kamsma2023}, which are all given, together with detailed expressions for $g_\infty(U)$, in \cref{sec:ginf}. Of particular importance is the characteristic time scale $\tau\approx5\,\mathrm{ms}$ of our parameter set,  which is the memory retention time of the microfluidic memristor. We note, however, that the dynamic response of the conical channel is qualitatively independent of the details of the system parameters.

\begin{figure*}[t!]
  \centering
  \includegraphics[width=\linewidth]{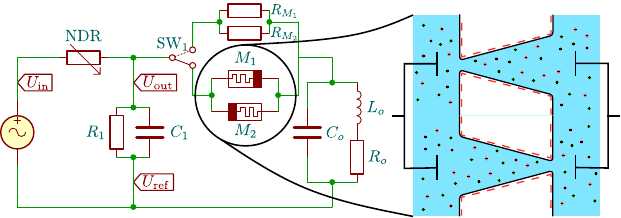}
  \caption{%
  Schematic of a single Memriki oscillator.
  The capacitor $C_o$, inductor $L_o$ and resistor $R_o$ together form a damped oscillator.
  The circuit is driven by the input voltage $U_\mathrm{in}$, measured relative to the reference voltage $U_\mathrm{ref}$.
  A negative differential resistance $\mathrm{NDR}$ introduces nonlinearity into the circuit.
  The output voltage $U_\mathrm{out}$ is measured,  stabilized by $C_1$, and slowly relaxed towards $U_\mathrm{ref}$ by $R_1$.
  Two iontronic memristors $M_1$ and $M_2$ are added between $U_\mathrm{out}$ and the oscillatory section to introduce further nonlinearity.
  With the switch SW$_1$ the memristors can be replaced with ordinary resistors.
  }
  \label{fig:shinriki_schem}
\end{figure*}%
We consider a sinusoidal voltage $U(t) = U_0\sin(2\pi ft)$ with fixed amplitude $U_0=1\V$  and several frequencies $f$ applied over the microfluidic channel to analyze its frequency-dependent dynamic response resulting from numerical solutions of Eq. (\ref{eq:ginf}). Focusing on the limit cycle in which all transients have decayed, we present in Fig. \ref{fig:channel} parametric plots for three different frequencies,  $(U(t),g(t))$ in (b) and $(U(t),I(t))$ in (c), where the current is defined as $I(t)=g(t)U(t)$. Depending on the driving frequency, the electric response of the channel falls into one of three regimes: (i)~At low frequencies $f \ll 1/\tau$,  exemplified here by $f=0.4\,\mathrm{Hz}$ (blue), the conductance $g(U(t))$ closely follows the steady-state function $g_\infty(U(t))$ in (b), as the driving voltage changes slowly enough for the concentration profile to  fully build up at all times. In this low-frequency regime, the electric current $I(t)$ in (c) shows strong rectification, approaching nearly full diodic behavior, with significantly higher current at negative voltages than at positive ones.
(ii)~At high frequencies $f \gg 1/\tau$,  exemplified here by $f=400\,\mathrm{Hz}$ (green), the sinusoidal voltage oscillates too rapidly for  the ionic concentration profile to develop, leaving it essentially spatially and temporally constant.
As a result, the conductance is essentially Ohmic with constant $g(t)\approx 1.07\,\mathrm{g_0}$ in (b) and $I(t)$ linear in $U(t)$ in (c).
One might have expected the conductance in this high-frequency regime to be exactly equal to $g_0$.  However, the nonlinearity of $g_\infty(U)$ results in a nonzero difference between the static conductance at the time-averaged voltage $g_0$ and the time-averaged  steady-state conductance for the time-dependent voltage, yielding $\langle g_\infty(U(t) \rangle\approx1.07\,\mathrm{g_0}$.
(iii)~At intermediate frequencies $f\sim 1/\tau$, exemplified here by $f=40\,\mathrm{Hz}$ (orange), the driving voltage varies on a timescale comparable to the memory timescale of the channel. As a result, hysteresis emerges
in both the conductance (b) and the current (c) since the ionic concentration profile can partially build up and break down but not fully equilibrate. This  leads to a lower channel conductance when the potential was recently positive and a higher conductance when it was recently negative.

In the remainder of this work, we measure all quantities relative to the steady-state conductance $g_0$ of the memristor at $0\V$. We use Volts ($\V$) and seconds ($\s$) as our units for potential and time, respectively. Specifically, resistance is expressed in $\RR$, current in $\A$, capacitance in $\CC$, and inductance in $\LL$.
We solve the equations of motion for all circuits in this paper using the circuit simulation software \texttt{ACME.jl} based on~\cite{ACME}.
For all simulations, we use a time step $\delta t=10^{-5}\s$, except for the calculations in Fig.~\ref{fig:channel} (b) and (c), where we use $\delta t=10^{-5}/f$, to ensure sufficient resolution for the high frequency cases.
The code for reproducing the simulations is provided in Ref.~\cite{code}.

\section{Characterization of a Single Memriki oscillator}
\begin{figure*}[htpb]
  \centering
  \includegraphics[width=\linewidth]{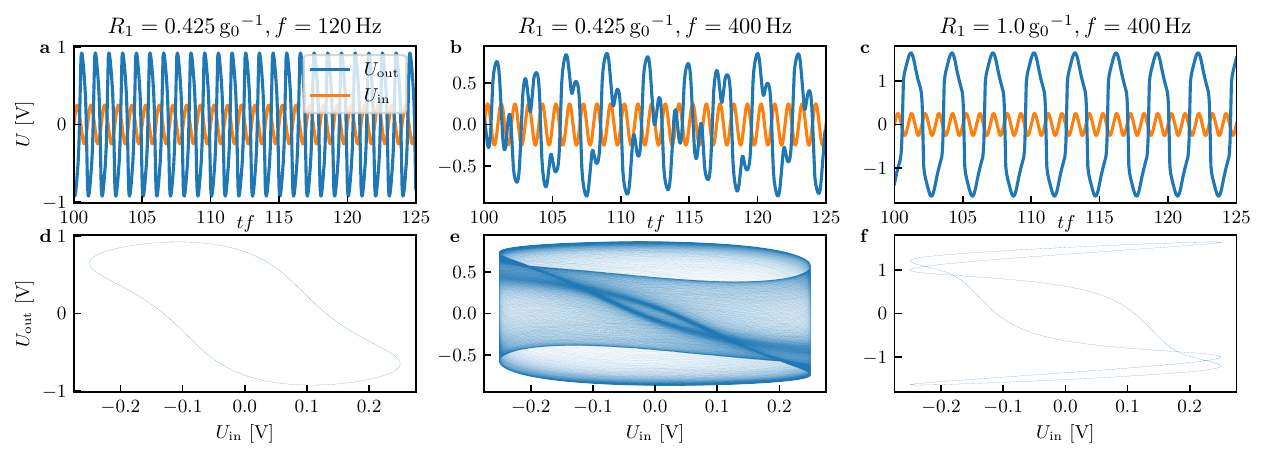}
  \caption{%
  Dynamics of the driven Memriki oscillator from Fig. \ref{fig:shinriki_schem} for three different combinations of resistor $R_1$ and driving frequency $f$, with all other circuit elements specified in the text. The top panels (a)-(c) show the time series of the input voltage $U_\mathrm{in}=0.25\sin(2\pi ft)\V$ (orange) and the output voltage $U_\mathrm{out}$ (blue) during 25 driving periods after 100 initial periods. The bottom panels (d)-(f) show the corresponding parametric plots of $U_\mathrm{out}$ versus $U_\mathrm{in}$ over 1000 driving periods, revealing stable periodic orbits in (d) and (f) separated by a chaotic, non-periodic regime in (e).
  Thus, in (a,d) the Memriki oscillator follows the input signal and oscillates at the driving frequency $f$, in (b,e) the response exhibits  chaotic behavior, while in (c,f), the response is subharmonic with a threefold period at an oscillating frequency $f/3$.
  }
  \label{fig:dynamics}
\end{figure*}
To leverage the volatile behavior of the memristors of Section II, we incorporate two of them into a nonlinear electrical circuit with an intrinsic oscillatory timescale similar to that of the memristors.
The  circuit design, shown in Fig.~\ref{fig:shinriki_schem}, is heavily inspired by the well-known Shinriki-oscillator~\cite{Shinriki1981},  albeit with a slightly modified topology and with iontronic memristors $M_1$ and $M_2$ rather than diodes.
We propose the name \emph{Memriki oscillator} for this circuit.
Later we will swap out the memristors for conventional resistors to study the role of the memristors. In order to allow for this swapping, we introduced the switch SW$_1$ into the circuit, to either connect the resistors or the memristors.
The circuit, driven by a sinusoidal source potential $U_{in}=U_0\sin(2\pi ft)$ with amplitude $U_0$ and frequency $f$, consists of a negative differential resistance $\mathrm{NDR}$, two anti-parallel memristors in series with a (weakly damped) LRC oscillator with eigen frequency $1/(2\pi\sqrt{L_oC_o})$, and a resistor $R_1$ and capacitor $C_1$ that are both parallel to the memristor-oscillator part. Based on the original work of Shinriki {\em et al}. ~\cite{Shinriki1981}, the
$\mathrm{NDR}$ is an active element with a characteristic function given by  $R(U)\,\mathrm{g_0} = (U/\V)^3-3U/\V$. This implies that it behaves as an active amplifier rather than  a passive resistor within the $\pm1\V$ voltage range around $0\V$. The parallel resistor and capacitor, with  $g_0R_1$  of order unity and $C_1=10^{-4}\CC$, respectively, can serve as tunable parameters that significantly alter  the dynamics of the circuit.
The electric specifications for the oscillatory part of the circuit read $C_o=10^{-3}\CC$, $L_o=10^{-3}\LL$ and $R_o=10^{-3}\RR$, ensuring that $R_o\ll R_1$. As a result,  the eigen period of the  weakly damped oscillator is approximately  $2\pi\sqrt{L_oC_o}\approx 6\,\mathrm{ms}$, which is comparable to the memory time of the two memristors, $\tau\approx5\,\text{ms}$.
While we describe and denote all circuit elements (except for the memristors) as conventional electrical components, we note that microfluidic counterparts exist for each of them~\cite{Stojanovic2019,GallardoHevia2022}. This  enables a purely microfluidic realization of this Memriki oscillator.

The dynamic response  of the Memriki oscillator varies dramatically depending on the precise values of the resistor $R_1$ and the driving frequency $f$. For three different combinations of $R_1$ and $f$ (and for $U_0=0.25\V$ and all other parameters fixed as in the text above) this parameter sensitivity of the response is illustrated in Fig.~\ref{fig:dynamics}. Here the top row (a)-(c) shows the time-dependence of the driving voltage $U_\mathrm{in}(t)=U_0\sin(2\pi ft)$ (orange) with the resulting output voltage $U_\mathrm{out}(t)$ (blue); the bottom row (d)-(e) shows the parametric plot $(U_\mathrm{in}(t),U_\mathrm{out}(t))$. Fig.~\ref{fig:dynamics} thus exhibits behaviors ranging from simple driven oscillations at $R_1=0.425 g_0^{-1}$ and $f=120$ Hz in (a,d), to chaotic dynamics  by increasing the frequency to $f = 400$ Hz in (b,e), and finally to subharmonic oscillations with frequency $f/3$ by subsequently increasing the resistance to $R_1 = 1.0 g_0^{-1}$ while maintaining $f=400$ Hz in (c,f).

To quantify the complexity of the output voltage $U\ind{out}(t)$ in greater detail, we count the number of distinct  output voltage values at discrete times $t=n/f$, where  $n\in\mathbb{N}$, i.e.\ at  time points corresponding to integer multiples of the input signal's periodicity. The number of recurrences of distinct output values is defined as  $N\ind{rec}=\left|\{U\ind{out}(n/f)\quad \text{for } n\in \mathbb{N}\}\right|$, where $|\cdot |$ denotes  the number of elements in the set. In practice, we consider two output values to be the same if their (absolute) difference is smaller than a typical  tolerance, $\epsilon=0.01\V$, to account for finite numerical precision. For example, this definition yields  $N\ind{rec}=1$  for the periodic oscillation in Fig. \ref{fig:dynamics}(a) and  $N\ind{rec}=3$ for the subharmonic oscillation in Fig. \ref{fig:dynamics}(c). Similarly, $N\ind{rec}$ is a small integer for  periodic output voltages $U\ind{out}(t)$ that are commensurate with the input signal $U\ind{in}(t)$. However, for incommensurate or chaotic output voltages,  the number of distinct output values diverges, i.e.\ $N\ind{rec}\rightarrow\infty$. To ensure that the computation of $N\ind{rec}$ remains numerically tractable, we first let the system equilibrate for 100 periods of the input signal and then analyze the next 100 periods. Thus,\ we restrict the analysis to $n\in [101,200]$, which bounds  $N\ind{rec}\in[1,100]$. This approach gives,  for instance, $N\ind{rec}=61$ for the case of Fig.~ \ref{fig:dynamics}(b) that we would deem to be chaotic for all intents and purposes. In the analysis and figures below,  we further simplify the presentation and only distinguish $N\ind{rec}\in[1,20]$.

\begin{figure}[htbp]
  \centering
  \includegraphics[width=\linewidth]{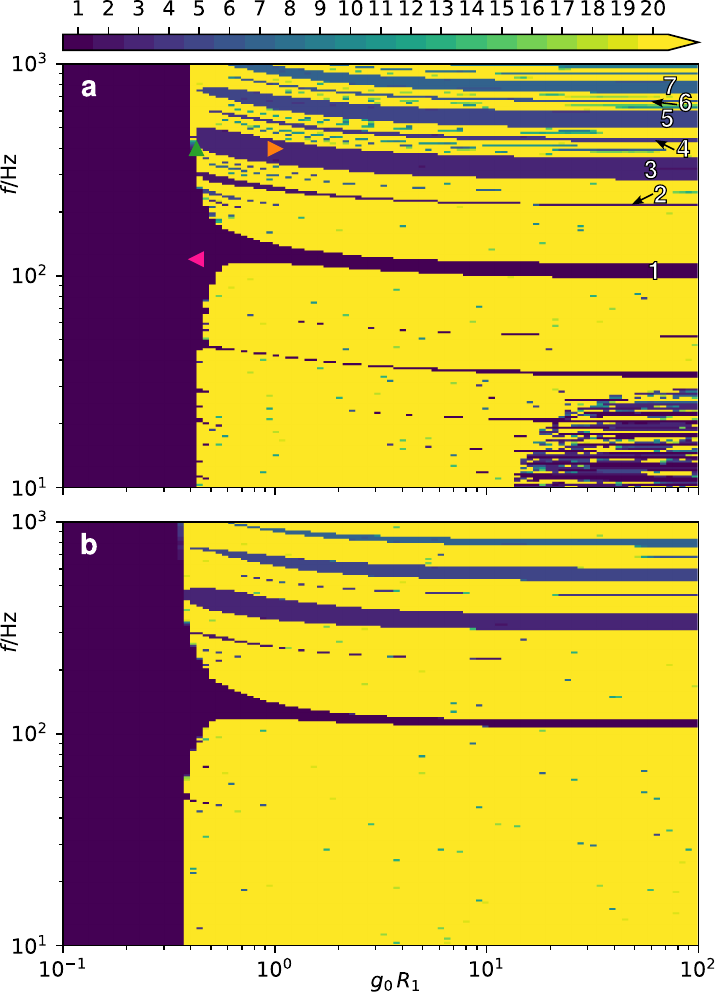}
  \caption{
    Heat map of the number of recurrences $N\ind{rec}$ of distinct output voltage values at times corresponding to integer periods of the input (see text), as a function of the driving frequency $f$ and the resistance of $R_1$ in the circuit shown in Fig. \ref{fig:shinriki_schem}. Results are shown for the two positions of the switch SW$_1$, which  couples either (a) the memristors or (b) the Ohmic resistors to the RCL oscillator. All other parameter values are provided in the text.
    Numbers in (a) indicate $N\ind{rec}$ for the corresponding bands.
    Marked points show the state-points of~\cref{fig:dynamics}. Pink left pointing triangle for~\cref{fig:dynamics} (a), green upwards pointing triangle for (b) and orange right pointing triangle for (c).
  }
  \label{fig:phases}
\end{figure}

In \cref{fig:phases}(a) we show the heatmap of $N\ind{rec}$ for the circuit of Fig.~\ref{fig:shinriki_schem} as a function of $R_1$ and $f$, with the switch SW$_1$ in the ``memristor state'' and with all other parameters given in the text. We can clearly distinguish a (fully purple) low-resistor regime $g_0R_1<0.4$ where the oscillator is completely stable with $N\ind{rec}=1$, regardless the driving frequency $f$. The high-resistor regime $g_0R_1>0.4$ shows a richer palette of behavior with large chaotic regimes (yellow) where $N\ind{rec} \geq 20$ and substantial stable bands (purple) with $N\ind{rec}=1$ dispersed by tiny intermediate regimes (green). Several of the stable (purple) bands extend deep into the high-resistor regime, especially at high frequencies $f>100\Hz$. Also, for $g_0R_1>10$ and low frequencies $f<10^{1.5}\Hz$, a regime appears where chaos and stability are closely inter-dispersed.
Furthermore, we  note that the stable (purple) bands with subharmonic response are significantly wider when the response frequency is an odd multiple of the input frequency as compared to an even multiple.

In order to analyze (and appreciate) the role of the memristors, we show the heat map of $N\ind{rec}$ again in Fig.~\ref{fig:phases}(b), however now with the switch SW$_1$ set to the ``resistor state'' with two parallel Ohmic resistors replacing the two parallel memristors. The value $R_{M_{1,2}}=0.25\RR$ that we chose for the two resistors
might seem a bit low at first sight, however this is justified because one of the two memristors in (a) is always in the low-resistance state due to their anti-parallel wiring. Comparing the Ohmic case of Fig. \ref{fig:phases}(b) with the memristor case of (a) reveals a similar low-resistance stability regime, however in the predominantly chaotic (yellow) regime at higher resistances much fewer stable (purple) bands appear and their (purple-yellow) interfaces appear to be sharper in (b) indicating fewer modes with intermediate harmonic periods of several driving periods. Moreover, the finely dispersed stable-and-chaotic regime found at low $f$ and high $R_1$ in (a) has disappeared altogether in (b). In other words, the memristors increase the regime of predictability in the predominantly chaotic high-resistance regime, and therefore contribute to a circuit being at the edge of chaos. For this reason, we focus on circuits of Memriki oscillators below.

\section{Coupling Memriki oscillators}
\begin{figure*}[htpb]
  \centering
  \includegraphics[width=\linewidth]{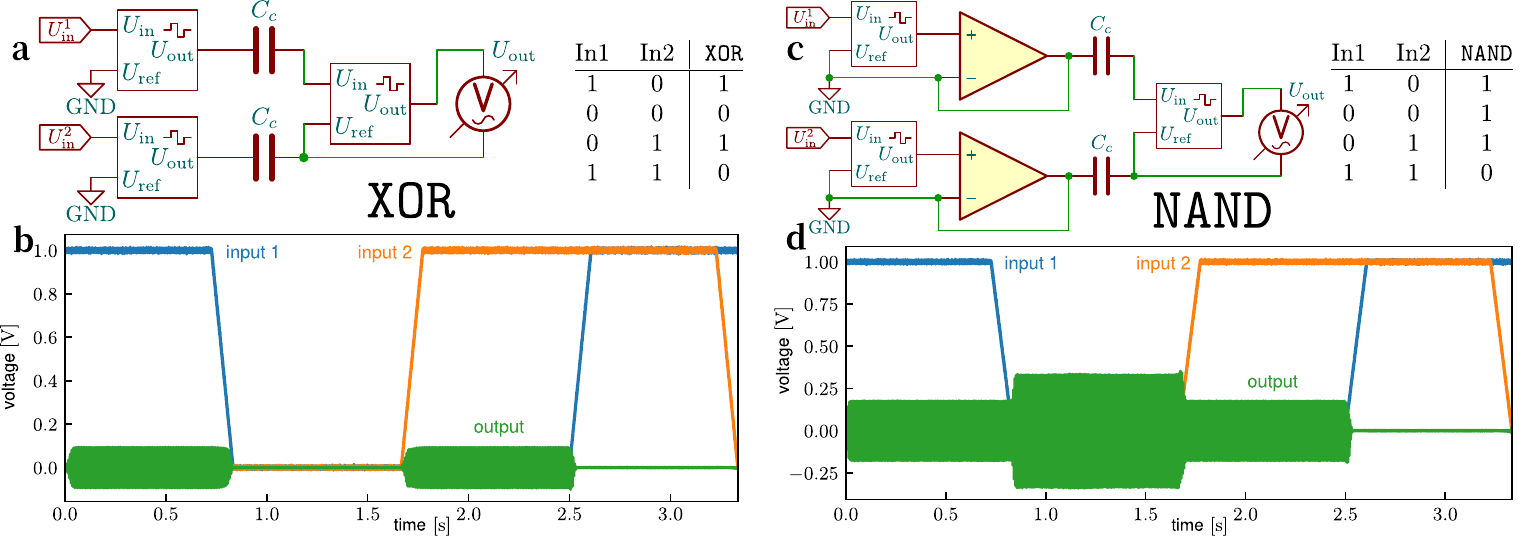}
  \caption{%
    Circuit diagrams of a \texttt{XOR}-pre-gate (a) and a \texttt{NAND}-pre-gate (c) constructed by  combining three Memriki oscillators from  \cref{fig:shinriki_schem} (represented here by the three boxes), either  coupled solely via capacitors (a) or with the addition of two  operational amplifiers (c).
    Time traces of the two input voltages (blue and orange) representing the four consecutive binary states 10, 00, 01, 11. The resulting quasi-binary  output voltage (green), which is either quiescent or oscillating, is shown in (b) for circuit (a), revealing  \texttt{XOR} behavior, and in (d) for circuit (c), exhibiting   \texttt{NAND} behavior.
    The corresponding truth tables are shown next to  the circuit diagrams.
    }
  \label{fig:shinriki_schem_xor}
\end{figure*}

The most fundamental computations involve combining two inputs to produce a single output.
To achieve this, we couple three Memriki oscillators, represented by the three ``boxes'' in the circuit diagrams of \cref{fig:shinriki_schem_xor} (a) and (c). Two oscillators serve as inputs, while the third acts as an output.
In \cref{fig:shinriki_schem_xor}(a), we see that the outputs of the two input oscillators  are capacitively connected to the input of the output oscillator,  which prevents any DC current from flowing from the inputs to the output.
When the circuit of \cref{fig:shinriki_schem_xor}(a) is driven by input voltages $0$ and $1\,\V$, representing low and high logic levels, along with Gaussian noise with a standard deviation of $0.003\V$, the output of the circuit turns out to correspond to a logic \texttt{XOR}-like gate. For $R_1=0.4\RR$, this is shown in the voltage time trace of \cref{fig:shinriki_schem_xor}(b), with blue and orange representing the four consecutive input combinations (10, 00, 01, 11) and green the output.  That is, the circuit oscillates when the two inputs differ and remains quiescent when they  are the same. This \texttt{XOR}-type behavior can be interpreted as performing a nonlinear classification task, one of the most fundamental classification tasks in machine learning~\cite{minsky2017perceptrons}.
For reasons that will become clear in the next section we refer to this as an \texttt{XOR}-pre-gate.

The \texttt{NAND}-gate is the primary objective when building logic gates, as all other gates can be constructed from combinations of \texttt{NAND}.
We realize a \texttt{NAND}-pre-gate by inserting two additional operational amplifiers into the circuit as shown in \cref{fig:shinriki_schem_xor}(c).
These amplifiers function as impedance changers,  ensuring  sustained oscillations in the third oscillator when both inputs are low.
The resulting output voltage, shown in green in \cref{fig:shinriki_schem_xor}(d) for the same four input combinations (blue and orange) as in (b), oscillates for the  input states 10, 00, and 01, but remains quiescent for input 11. Therefore, we classify this circuit as a \texttt{NAND}-pre-gate.

\begin{figure}[htbp]
  \centering
  \includegraphics[width=\linewidth]{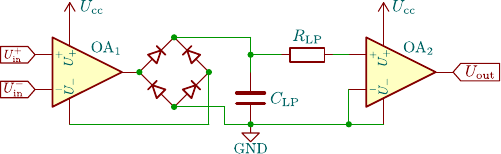}
  \caption{%
    Coupler for converting an oscillatory signal into constant logic levels.
    The oscillatory input voltage difference between $U_\mathrm{in}^+$ and $U_\mathrm{in}^-$ is first amplified by the left operational amplifier OA$_1$. This amplified signal is then rectified by a full-wave bridge rectifier, consisting  of four diodes.
    The rectified signal is subsequently smoothed by a low-pass filter consisting of $R_\mathrm{LP}$ and $C_\mathrm{LP}$ and finally amplified to the desired logic level by the right operational amplifier OA$_2$.
    The operational amplifiers in this circuit are modeled by \cref{eq:macak}, with a maximum voltage of  $U_+=U\ind{cc}=1\V$.
  }
  \label{fig:coupler}
\end{figure}

To use the output of one logic gate as the input for another, the output must be compatible with the input requirements, in this case static signals of $0\V$ and $1\V$.
To achieve a proper \texttt{NAND} gate, we must transform the oscillatory output of our \texttt{NAND}-pre-gate
into a constant output of $1\V$ while its quiescent (0) output for the 11 input remains unaffected. This transformation is  achieved by passing the output voltage of our \texttt{NAND}-pre-gate through the coupler shown in the circuit diagram of \cref{fig:coupler}.
In this coupler, the output voltage first passes through the  operational amplifier OA$_1$, which strengthens the signal and prevents drawing any current from the input gates.
The amplified signal is then passed through a full-wave bridge rectifier composed of four diodes, followed by time-averaging with a low-pass filter with time constant $R_{LP}C_{LP}=0.5\,\mathrm{ms}$. Finally, the signal is  amplified by OA$_2$ to achieve the desired $1\V$ logic level.

The two operational amplifiers used in the coupler convert the input voltage $U_\mathrm{in}$ into an output voltage given by
\begin{equation}
  U_{\mathrm{out}}=\Delta U \tanh \left( \frac{G U_\mathrm{in}}{\Delta U} \right),
  \label{eq:macak}
\end{equation}
where $G$ is the gain factor (for small inputs),  set to $G=15$ for $\mathrm{OA}_1$ and to $G=35$ for $\mathrm{OA}_2$. The parameter $\Delta U=U_+-U_-=1.0\V$ defines the range of the amplifier, allowing  amplification up to $\pm 1\V$.
The low-pass RC filter consists of a resistance  $R_\mathrm{LP}=10^{-2}\RR$ and a capacitance  $C_\mathrm{LP}=5\times 10^{-2}\CC$, while the diodes are modeled with a saturation current of $0.3\A$ and an emission factor $\eta=1$~\cite{Shockley1949}.

With the addition of this coupler circuit, the \texttt{NAND}-pre-gate becomes a true \texttt{NAND} gate.

\section{Combining gates}
\begin{figure}[htbp]
  \centering
  \includegraphics[width=\linewidth]{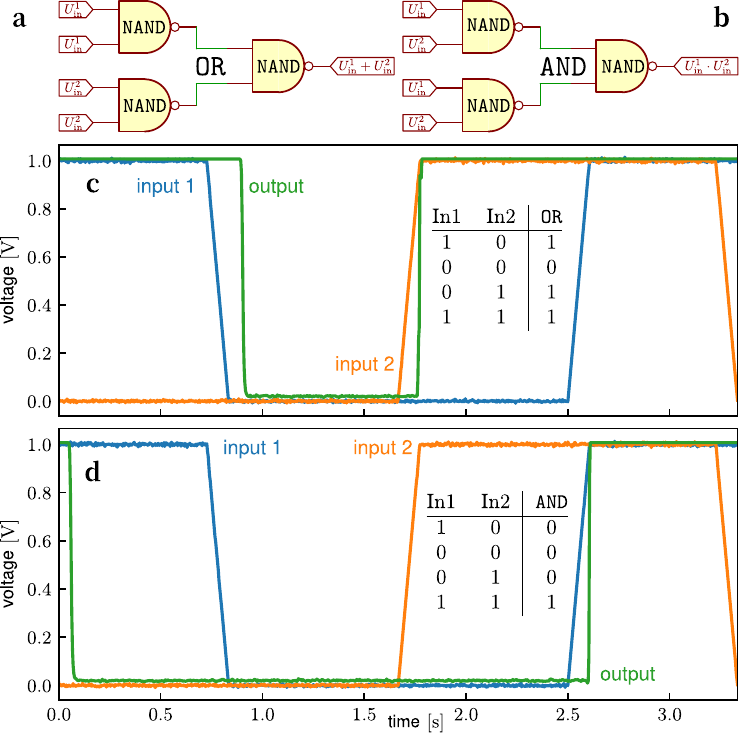}
  \caption{%
  Three combined \texttt{NAND} gates (see text) form (a) an \texttt{OR} gate and (b) an \texttt{AND} gate. Time traces of the input
  voltages (blue and orange), representing four consecutive binary states, and the corresponding output voltage (green) for (c) the \texttt{OR} gate and (d) the \texttt{AND} gate, along  with the corresponding truth tables.
  The delay in the switching of the output signal after the inputs have changed originates from the low-pass filter in the coupling circuit in ~\cref{fig:coupler}.
  }
  \label{fig:ANDOR}
\end{figure}

The \texttt{NAND} gate is well known to be functionally complete, which means that any other logic gate can be constructed from it.
In particular, an \texttt{OR} gate can be constructed using  three \texttt{NAND} gates, as shown in \cref{fig:ANDOR}(a).
The Boolean logic for the \texttt{OR} gate follows $ \nd{\nd{U_\mathrm{in}^1}{U_\mathrm{in}^1}}{\nd{U_\mathrm{in}^2}{U_\mathrm{in}^2}} = \nd{\overline{U_\mathrm{in}^1}}{\overline{U_\mathrm{in}^2}} = \overline{\overline{U_\mathrm{in}^1}} + \overline{\overline{U_\mathrm{in}^2}} = U_\mathrm{in}^1+U_\mathrm{in}^2$, where $\cdot$ denotes the logic \texttt{AND}, $+$ the logic \texttt{OR}, and an overline indicates negation. In our circuits, each input signal  $U_\mathrm{in}^1$ and $U_\mathrm{in}^2$ is passed  through the combined gates twice to ensure independent noise sources for all our Memriki gates.
If identical signals were used, the differential input to the final gate would cancel out, preventing the oscillations from self-exciting.
To further aid self-excitation, we also add Gaussian noise with a standard deviation of $0.003\V$ to the inputs of the last gate.

The voltage time trace  in \cref{fig:ANDOR}(c) displays the output (green) of the \texttt{OR} gate for the same four consecutive input signals (blue and orange) used  previously. This  confirms that the gate functions correctly as a  logic \texttt{OR} gate, processing  weakly noisy static inputs. The   response time of the gate is approximately  $0.1\,\mathrm{s}$ for relatively slowly varying input signals.

As is well known, the circuit for the \texttt{OR} gate can be reused to realize an \texttt{AND} gate; only the inputs need to be rearranged such that  both  input gates now receive $U_\mathrm{in}^1$ and $U_\mathrm{in}^2$, as shown in \cref{fig:ANDOR}(b).
The logic for the \texttt{AND} gate is as follows: $\nd{\nd{U_\mathrm{in}^1}{U_\mathrm{in}^2}}{\nd{U_\mathrm{in}^1}{U_\mathrm{in}^2}} = \overline{\nd{U_\mathrm{in}^1}{U_\mathrm{in}^2}} + \overline{\nd{U_\mathrm{in}^1}{U_\mathrm{in}^2}} = U_\mathrm{in}^1 \cdot U_\mathrm{in}^2$. The resulting input-output relation for this \texttt{AND} gate is shown in \cref{fig:ANDOR}(d), confirming  that it functions correctly as an \texttt{AND} gate. The dynamic response to slowly varying inputs is comparable to that of the \texttt{OR} gate.

To test the repeatability and reproducibility of the gates we need to verify whether they perform as expected over many cycles of the (weakly noisy) input and output sequences.
For our continuous system, with output voltages spanning the full range between zero and one volt when the input signal switches, we  need to define boundaries separating high- and low-level signals during specific time intervals. Here we set the time interval as the duration between the switching of the input  signals and consider an output larger than $0.66\V$ as a digital 1 and below $0.33\V$ a digital 0.
To evaluate the pre-gates, we connect them to the coupling circuit as described above to ensure they function as proper gates.
Using these definitions, we tested 100 cycles of all possible input combinations and obtained the following accuracies. Individual \texttt{XOR} gates and \texttt{NAND} gates operate with an accuracy of $\geq 99\%$,  the combined \texttt{OR} gate  achieves also $\geq 99\%$ accuracy, and the combined \texttt{AND} gate reaches an accuracy of $\geq 95\%$.

\section{Conclusion}
We designed microfluidic iontronic circuits, primarily combining volatile memristors with Shinriki-like oscillators, that can perform  non-linear \texttt{XOR} classification  and act as functional complete \texttt{NAND} logic gates.
We showed that these \texttt{NAND} gates are robust and can be combined to construct other logic gates, especially \texttt{OR} and \texttt{AND}. A key feature of these circuits is their exploitation of  noise, here introduced at a level of $3\,\mathrm{mV}$ relative to  a $1\,\mathrm{V}$ signal scale. This noise level enables the self-excitation of  oscillators from a quiescent (0) to an oscillating (1) state,  a feature that stems from the intrinsic non-linearity of the ``Memriki'' oscillators from which our circuits were composed. We hope  our findings will inspire future experimental exploration  of memristive iontronic devices. Although  we have not yet assessed the energy efficiency of our circuits and logic gates in  detail, we do not expect them to outperform  present-day silicon-based  technologies in this regard. In this sense, this work serves as a proof of concept for a potentially versatile iontronic computing platform, operating with  the same aqueous electrolyte environment as mammalian brains.

\section{Methods}
\label{sec:ginf}
To evaluate the  steady-state conductance of the conical channels, we use the parameters and expressions from~\cite{Kamsma2023} for \cref{eq:ginf}. For a conical channel of length $L=10\,\mathrm{\mu m}$, base radius $R_b=200\,\mathrm{nm}$, tip radius $R_t=50\,\mathrm{nm}$, and a radius profile $R(x)=R_tx/L+R_b(1-x/L)$ for $x\in[0,L]$ running from base to tip, the static conductance $g_\infty(U)$ is given by
\begin{eqnarray}
  \frac{g_\infty(U)}{g_0}& = &\nonumber \\
  & & \hspace{-5mm}1+\Delta g\int_0^L\left( \frac{xR_t}{LR(x)}-\frac{\e
    ^{ \frac{x\Pe R_t^2}{LR_bR(x)}} -1}{\e^ { \Pe \frac{R_t}{R_b} } -1}\right) \frac{\dx}{L}.
\end{eqnarray}
Here, $g_0\approx4.2\,\mathrm{pS}$ is the zero-field conductance, $\Delta g \approx -3.59$ is the asymmetry parameter, and $\Pe$ is the Peclet number, which varies  linearly with the applied voltage $U$ as  $\Pe(U)\approx16U/\mathrm{V}$ for our parameter set (exact expressions provided below).  Using the short-hand notation $\Delta R=R_b-R_t$, and performing the integral, one obtains the explicit static channel conductance
\begin{gather}
  g_\infty(U) = g_0 \left(  1 + \Delta g\frac{R_t}{\Delta R} F(U) \right)\\
  F(U) =     \frac{\e^{\frac{-\Pe  R_t^2}{\Delta R R_b}}A(U)}{\e^{\Pe \frac{R_t}{R_b}} - 1} -\left(\frac{R_b}{\Delta R} \ln\frac{R_t}{R_b} + 1\right) \\
  A(U) =   \e^{\frac{\Pe R_t}{\Delta R}} - \frac{R_b}{R_t}
  \e^{\frac{\Pe R_t^2}{\Delta R R_b}} + \frac{\Delta R}{R_t} + B(U) \\
  B(U) =  \frac{-\Pe R_t}{\Delta R} \left(\Ei\left(\frac{\Pe R_t}{\Delta R}\right) - \Ei\left(\frac{\Pe R_t^2}{\Delta R R_b}\right)\right),
\end{gather}
where $\Ei(x)=\int_{-\infty}^x\e^{-t}/t \dd t$ is the exponential integral function.

We stress that this explicit expression for $g_\infty(U)$ stems from an (approximate) analytic solution to the fully microscopic Poisson-Nernst-Planck-Stokes equations in the long-channel limit. The dependence of the conductance on $U$ is a direct consequence of  electro-osmotic flow, characterized by the dimensionless parameter $\Pe(U)$. This effect relies on the presence of  a nonzero surface charge $e\sigma$ (and hence a nonzero zeta potential $\psi$ on the channel walls), as becomes explicit by the analytic relation $\Pe(U) = \frac{-eR_b}{k_BT  R_t w} U$, where  $w=\frac{e D \eta}{k_BT \varepsilon \psi}\approx-9.6$ represents the ratio of ionic mobility $D/k_BT$ to osmotic mobility $\varepsilon\psi/(e\eta)$. Here  $D=1.75\,\mathrm{nm}^2/\mathrm{ns}$ is the diffusion coefficient of the monovalent cations and anions in water, $k_B$ denotes the Boltzmann constant, $T=293.15\,\mathrm{K}$ is the absolute temperature, and $\varepsilon=80.2\epsilon_0\approx7.1\times10^{-8}\,\mathrm{F/m}$ is the dielectric constant, and $\eta=1.01\,\mathrm{mPa s}$ is the shear viscosity of water.
We define the steady zero-field conductance as $g_0= \frac{ \pi R_tR_b}{L} \frac{2\rho_b e^2 D}{k_BT}$, where $\rho_b$ is the ionic bulk concentration, set  to $10\,\mathrm{mM}$.
The asymmetry parameter is given by
$\Delta g=\frac{-2w \Delta R}{R_b \mathrm{Du}}\approx-3.59$,
where the Dukhin number is $\mathrm{Du}=\sigma / (2\rho_b R_t)\approx-0.25$, with $e\sigma=-0.0015\,e/\mathrm{nm}^2$ the surface charge density on the channel walls. The corresponding   zeta potential is $\psi=\frac{2k_BT}{e} \mathrm{asinh}(2\pi \lambda_D \lambda_B \sigma)\approx-10\,\mathrm{mV}$, where  $\lambda_D=\sqrt{\varepsilon k_BT / (2e^2\rho_b)}\approx30\,\mathrm{nm}$ is the Debye length and $\lambda_B=e^2 / (4\pi \varepsilon k_BT)\approx0.71\,\mathrm{nm}$ is the Bjerrum length.

The characteristic time scale of this conical channel, which is the time scale to build up (or break down) the static voltage-dependent salt concentration profile $\rho_s(x;U)$ in the channel, was derived to be the diffusion time $\tau=L^2/12D$, valid in the long-channel limit $L\gg R_b$. For the present parameters this leads to $\tau\approx5\,\mathrm{ms}$.

\section{Data availability}
All data can be generated with the source-code provided in~\cite{code}.
The raw data is also available in Ref.~\cite{data}.

\section{Code availability}
The code to perform the simulations is provided in~\cite{code}.

\section{Competing Interests}
There are no competing interests.

\section{Author Contributions}
N.C.X.S designed the circuits, performed the simulations and analyzed the data. All authors discussed the results and contributed to the manuscript.

\section{Acknowledgements}
The authors thank Monica Conte and Tim Kamsma for helpful discussions.
N.C.X.S and M.D.\ acknowledge funding
from the European Research Council (ERC) under the
European Union’s Horizon 2020 research and innovation
program (Grant agreement No. ERC-2019-ADG 884902,
SoftML)
\bibliography{bib}

\begin{thebibliography}{49}%
\makeatletter
\providecommand \@ifxundefined [1]{%
 \@ifx{#1\undefined}
}%
\providecommand \@ifnum [1]{%
 \ifnum #1\expandafter \@firstoftwo
 \else \expandafter \@secondoftwo
 \fi
}%
\providecommand \@ifx [1]{%
 \ifx #1\expandafter \@firstoftwo
 \else \expandafter \@secondoftwo
 \fi
}%
\providecommand \natexlab [1]{#1}%
\providecommand \enquote  [1]{``#1''}%
\providecommand \bibnamefont  [1]{#1}%
\providecommand \bibfnamefont [1]{#1}%
\providecommand \citenamefont [1]{#1}%
\providecommand \href@noop [0]{\@secondoftwo}%
\providecommand \href [0]{\begingroup \@sanitize@url \@href}%
\providecommand \@href[1]{\@@startlink{#1}\@@href}%
\providecommand \@@href[1]{\endgroup#1\@@endlink}%
\providecommand \@sanitize@url [0]{\catcode `\\12\catcode `\$12\catcode
  `\&12\catcode `\#12\catcode `\^12\catcode `\_12\catcode `\%12\relax}%
\providecommand \@@startlink[1]{}%
\providecommand \@@endlink[0]{}%
\providecommand \url  [0]{\begingroup\@sanitize@url \@url }%
\providecommand \@url [1]{\endgroup\@href {#1}{\urlprefix }}%
\providecommand \urlprefix  [0]{URL }%
\providecommand \Eprint [0]{\href }%
\providecommand \doibase [0]{https://doi.org/}%
\providecommand \selectlanguage [0]{\@gobble}%
\providecommand \bibinfo  [0]{\@secondoftwo}%
\providecommand \bibfield  [0]{\@secondoftwo}%
\providecommand \translation [1]{[#1]}%
\providecommand \BibitemOpen [0]{}%
\providecommand \bibitemStop [0]{}%
\providecommand \bibitemNoStop [0]{.\EOS\space}%
\providecommand \EOS [0]{\spacefactor3000\relax}%
\providecommand \BibitemShut  [1]{\csname bibitem#1\endcsname}%
\let\auto@bib@innerbib\@empty
\bibitem [{\citenamefont {Chabal}(2001)}]{Chabal2001}%
  \BibitemOpen
  \bibinfo {editor} {\bibfnamefont {Y.~J.}\ \bibnamefont {Chabal}},\ ed.,\
  \href@noop {} {\emph {\bibinfo {title} {Fundamental aspects of silicon
  oxidation}}},\ Physics and astronomy online library\ (\bibinfo  {publisher}
  {Springer},\ \bibinfo {address} {Berlin [u.a.]},\ \bibinfo {year}
  {2001})\BibitemShut {NoStop}%
\bibitem [{\citenamefont {Burr}\ \emph {et~al.}(2016)\citenamefont {Burr},
  \citenamefont {Shelby}, \citenamefont {Sebastian}, \citenamefont {Kim},
  \citenamefont {Kim}, \citenamefont {Sidler}, \citenamefont {Virwani},
  \citenamefont {Ishii}, \citenamefont {Narayanan}, \citenamefont {Fumarola},
  \citenamefont {Sanches}, \citenamefont {Boybat}, \citenamefont {Le~Gallo},
  \citenamefont {Moon}, \citenamefont {Woo}, \citenamefont {Hwang},\ and\
  \citenamefont {Leblebici}}]{Burr2016}%
  \BibitemOpen
  \bibfield  {author} {\bibinfo {author} {\bibfnamefont {G.~W.}\ \bibnamefont
  {Burr}}, \bibinfo {author} {\bibfnamefont {R.~M.}\ \bibnamefont {Shelby}},
  \bibinfo {author} {\bibfnamefont {A.}~\bibnamefont {Sebastian}}, \bibinfo
  {author} {\bibfnamefont {S.}~\bibnamefont {Kim}}, \bibinfo {author}
  {\bibfnamefont {S.}~\bibnamefont {Kim}}, \bibinfo {author} {\bibfnamefont
  {S.}~\bibnamefont {Sidler}}, \bibinfo {author} {\bibfnamefont
  {K.}~\bibnamefont {Virwani}}, \bibinfo {author} {\bibfnamefont
  {M.}~\bibnamefont {Ishii}}, \bibinfo {author} {\bibfnamefont
  {P.}~\bibnamefont {Narayanan}}, \bibinfo {author} {\bibfnamefont
  {A.}~\bibnamefont {Fumarola}}, \bibinfo {author} {\bibfnamefont {L.~L.}\
  \bibnamefont {Sanches}}, \bibinfo {author} {\bibfnamefont {I.}~\bibnamefont
  {Boybat}}, \bibinfo {author} {\bibfnamefont {M.}~\bibnamefont {Le~Gallo}},
  \bibinfo {author} {\bibfnamefont {K.}~\bibnamefont {Moon}}, \bibinfo {author}
  {\bibfnamefont {J.}~\bibnamefont {Woo}}, \bibinfo {author} {\bibfnamefont
  {H.}~\bibnamefont {Hwang}},\ and\ \bibinfo {author} {\bibfnamefont
  {Y.}~\bibnamefont {Leblebici}},\ }\bibfield  {title} {\bibinfo {title}
  {Neuromorphic computing using non-volatile memory},\ }\href
  {https://doi.org/10.1080/23746149.2016.1259585} {\bibfield  {journal}
  {\bibinfo  {journal} {Advances in Physics: X}\ }\textbf {\bibinfo {volume}
  {2}},\ \bibinfo {pages} {89} (\bibinfo {year} {2016})}\BibitemShut {NoStop}%
\bibitem [{\citenamefont {Rajendran}\ \emph {et~al.}(2019)\citenamefont
  {Rajendran}, \citenamefont {Sebastian}, \citenamefont {Schmuker},
  \citenamefont {Srinivasa},\ and\ \citenamefont
  {Eleftheriou}}]{Rajendran2019}%
  \BibitemOpen
  \bibfield  {author} {\bibinfo {author} {\bibfnamefont {B.}~\bibnamefont
  {Rajendran}}, \bibinfo {author} {\bibfnamefont {A.}~\bibnamefont
  {Sebastian}}, \bibinfo {author} {\bibfnamefont {M.}~\bibnamefont {Schmuker}},
  \bibinfo {author} {\bibfnamefont {N.}~\bibnamefont {Srinivasa}},\ and\
  \bibinfo {author} {\bibfnamefont {E.}~\bibnamefont {Eleftheriou}},\
  }\bibfield  {title} {\bibinfo {title} {Low-power neuromorphic hardware for
  signal processing applications: A review of architectural and system-level
  design approaches},\ }\href {https://doi.org/10.1109/msp.2019.2933719}
  {\bibfield  {journal} {\bibinfo  {journal} {IEEE Signal Processing Magazine}\
  }\textbf {\bibinfo {volume} {36}},\ \bibinfo {pages} {97} (\bibinfo {year}
  {2019})}\BibitemShut {NoStop}%
\bibitem [{\citenamefont {Tripp}\ \emph {et~al.}(2024)\citenamefont {Tripp},
  \citenamefont {Perr-Sauer}, \citenamefont {Gafur}, \citenamefont {Nag},
  \citenamefont {Purkayastha}, \citenamefont {Zisman},\ and\ \citenamefont
  {Bensen}}]{tripp2024}%
  \BibitemOpen
  \bibfield  {author} {\bibinfo {author} {\bibfnamefont {C.~E.}\ \bibnamefont
  {Tripp}}, \bibinfo {author} {\bibfnamefont {J.}~\bibnamefont {Perr-Sauer}},
  \bibinfo {author} {\bibfnamefont {J.}~\bibnamefont {Gafur}}, \bibinfo
  {author} {\bibfnamefont {A.}~\bibnamefont {Nag}}, \bibinfo {author}
  {\bibfnamefont {A.}~\bibnamefont {Purkayastha}}, \bibinfo {author}
  {\bibfnamefont {S.}~\bibnamefont {Zisman}},\ and\ \bibinfo {author}
  {\bibfnamefont {E.~A.}\ \bibnamefont {Bensen}},\ }\href
  {https://arxiv.org/abs/2403.08151} {\bibinfo {title} {Measuring the energy
  consumption and efficiency of deep neural networks: An empirical analysis and
  design recommendations}} (\bibinfo {year} {2024}),\ \Eprint
  {https://arxiv.org/abs/2403.08151} {arXiv:2403.08151 [cs.LG]} \BibitemShut
  {NoStop}%
\bibitem [{\citenamefont {Baulin}\ \emph {et~al.}(2025)\citenamefont {Baulin},
  \citenamefont {Giacometti}, \citenamefont {Fedosov}, \citenamefont {Ebbens},
  \citenamefont {Varela-Rosales}, \citenamefont {Feliu}, \citenamefont
  {Chowdhury}, \citenamefont {Hu}, \citenamefont {Füchslin}, \citenamefont
  {Dijkstra}, \citenamefont {Mussel}, \citenamefont {van Roij}, \citenamefont
  {Xie}, \citenamefont {Tzanov}, \citenamefont {Zu}, \citenamefont
  {Hidalgo-Caballero}, \citenamefont {Yuan}, \citenamefont {Cocconi},
  \citenamefont {Ghim}, \citenamefont {Cottin-Bizonne}, \citenamefont {Miguel},
  \citenamefont {Esplandiu}, \citenamefont {Simmchen}, \citenamefont {Parak},
  \citenamefont {Werner}, \citenamefont {Gompper},\ and\ \citenamefont
  {Hanczyc}}]{Baulin2025}%
  \BibitemOpen
  \bibfield  {author} {\bibinfo {author} {\bibfnamefont {V.~A.}\ \bibnamefont
  {Baulin}}, \bibinfo {author} {\bibfnamefont {A.}~\bibnamefont {Giacometti}},
  \bibinfo {author} {\bibfnamefont {D.~A.}\ \bibnamefont {Fedosov}}, \bibinfo
  {author} {\bibfnamefont {S.}~\bibnamefont {Ebbens}}, \bibinfo {author}
  {\bibfnamefont {N.~R.}\ \bibnamefont {Varela-Rosales}}, \bibinfo {author}
  {\bibfnamefont {N.}~\bibnamefont {Feliu}}, \bibinfo {author} {\bibfnamefont
  {M.}~\bibnamefont {Chowdhury}}, \bibinfo {author} {\bibfnamefont
  {M.}~\bibnamefont {Hu}}, \bibinfo {author} {\bibfnamefont {R.}~\bibnamefont
  {Füchslin}}, \bibinfo {author} {\bibfnamefont {M.}~\bibnamefont {Dijkstra}},
  \bibinfo {author} {\bibfnamefont {M.}~\bibnamefont {Mussel}}, \bibinfo
  {author} {\bibfnamefont {R.}~\bibnamefont {van Roij}}, \bibinfo {author}
  {\bibfnamefont {D.}~\bibnamefont {Xie}}, \bibinfo {author} {\bibfnamefont
  {V.}~\bibnamefont {Tzanov}}, \bibinfo {author} {\bibfnamefont
  {M.}~\bibnamefont {Zu}}, \bibinfo {author} {\bibfnamefont {S.}~\bibnamefont
  {Hidalgo-Caballero}}, \bibinfo {author} {\bibfnamefont {Y.}~\bibnamefont
  {Yuan}}, \bibinfo {author} {\bibfnamefont {L.}~\bibnamefont {Cocconi}},
  \bibinfo {author} {\bibfnamefont {C.-M.}\ \bibnamefont {Ghim}}, \bibinfo
  {author} {\bibfnamefont {C.}~\bibnamefont {Cottin-Bizonne}}, \bibinfo
  {author} {\bibfnamefont {M.~C.}\ \bibnamefont {Miguel}}, \bibinfo {author}
  {\bibfnamefont {M.~J.}\ \bibnamefont {Esplandiu}}, \bibinfo {author}
  {\bibfnamefont {J.}~\bibnamefont {Simmchen}}, \bibinfo {author}
  {\bibfnamefont {W.~J.}\ \bibnamefont {Parak}}, \bibinfo {author}
  {\bibfnamefont {M.}~\bibnamefont {Werner}}, \bibinfo {author} {\bibfnamefont
  {G.}~\bibnamefont {Gompper}},\ and\ \bibinfo {author} {\bibfnamefont {M.~M.}\
  \bibnamefont {Hanczyc}},\ }\bibfield  {title} {\bibinfo {title} {Intelligent
  soft matter: towards embodied intelligence},\ }\href
  {https://doi.org/10.1039/d5sm00174a} {\bibfield  {journal} {\bibinfo
  {journal} {Soft Matter}\ }\textbf {\bibinfo {volume} {21}},\ \bibinfo {pages}
  {4129} (\bibinfo {year} {2025})}\BibitemShut {NoStop}%
\bibitem [{\citenamefont {Marković}\ \emph {et~al.}(2020)\citenamefont
  {Marković}, \citenamefont {Mizrahi}, \citenamefont {Querlioz},\ and\
  \citenamefont {Grollier}}]{Markovic2020}%
  \BibitemOpen
  \bibfield  {author} {\bibinfo {author} {\bibfnamefont {D.}~\bibnamefont
  {Marković}}, \bibinfo {author} {\bibfnamefont {A.}~\bibnamefont {Mizrahi}},
  \bibinfo {author} {\bibfnamefont {D.}~\bibnamefont {Querlioz}},\ and\
  \bibinfo {author} {\bibfnamefont {J.}~\bibnamefont {Grollier}},\ }\bibfield
  {title} {\bibinfo {title} {Physics for neuromorphic computing},\ }\href
  {https://doi.org/10.1038/s42254-020-0208-2} {\bibfield  {journal} {\bibinfo
  {journal} {Nature Reviews Physics}\ }\textbf {\bibinfo {volume} {2}},\
  \bibinfo {pages} {499} (\bibinfo {year} {2020})}\BibitemShut {NoStop}%
\bibitem [{\citenamefont {Chua}(1971)}]{Chua1971}%
  \BibitemOpen
  \bibfield  {author} {\bibinfo {author} {\bibfnamefont {L.}~\bibnamefont
  {Chua}},\ }\bibfield  {title} {\bibinfo {title} {Memristor-the missing
  circuit element},\ }\href {https://doi.org/10.1109/tct.1971.1083337}
  {\bibfield  {journal} {\bibinfo  {journal} {IEEE Transactions on Circuit
  Theory}\ }\textbf {\bibinfo {volume} {18}},\ \bibinfo {pages} {507} (\bibinfo
  {year} {1971})}\BibitemShut {NoStop}%
\bibitem [{\citenamefont {Strukov}\ \emph {et~al.}(2008)\citenamefont
  {Strukov}, \citenamefont {Snider}, \citenamefont {Stewart},\ and\
  \citenamefont {Williams}}]{Strukov2008}%
  \BibitemOpen
  \bibfield  {author} {\bibinfo {author} {\bibfnamefont {D.~B.}\ \bibnamefont
  {Strukov}}, \bibinfo {author} {\bibfnamefont {G.~S.}\ \bibnamefont {Snider}},
  \bibinfo {author} {\bibfnamefont {D.~R.}\ \bibnamefont {Stewart}},\ and\
  \bibinfo {author} {\bibfnamefont {R.~S.}\ \bibnamefont {Williams}},\
  }\bibfield  {title} {\bibinfo {title} {The missing memristor found},\ }\href
  {https://doi.org/10.1038/nature06932} {\bibfield  {journal} {\bibinfo
  {journal} {Nature}\ }\textbf {\bibinfo {volume} {453}},\ \bibinfo {pages}
  {80} (\bibinfo {year} {2008})}\BibitemShut {NoStop}%
\bibitem [{\citenamefont {Xia}\ and\ \citenamefont
  {Yang}(2019)}]{xia2019memristive}%
  \BibitemOpen
  \bibfield  {author} {\bibinfo {author} {\bibfnamefont {Q.}~\bibnamefont
  {Xia}}\ and\ \bibinfo {author} {\bibfnamefont {J.~J.}\ \bibnamefont {Yang}},\
  }\bibfield  {title} {\bibinfo {title} {Memristive crossbar arrays for
  brain-inspired computing},\ }\href@noop {} {\bibfield  {journal} {\bibinfo
  {journal} {Nature materials}\ }\textbf {\bibinfo {volume} {18}},\ \bibinfo
  {pages} {309} (\bibinfo {year} {2019})}\BibitemShut {NoStop}%
\bibitem [{\citenamefont {Jeon}\ \emph {et~al.}(2024)\citenamefont {Jeon},
  \citenamefont {Ryu}, \citenamefont {Im}, \citenamefont {Seo}, \citenamefont
  {Eom}, \citenamefont {Ju}, \citenamefont {Yang}, \citenamefont {Jeong},\ and\
  \citenamefont {Kim}}]{jeon2024purely}%
  \BibitemOpen
  \bibfield  {author} {\bibinfo {author} {\bibfnamefont {K.}~\bibnamefont
  {Jeon}}, \bibinfo {author} {\bibfnamefont {J.~J.}\ \bibnamefont {Ryu}},
  \bibinfo {author} {\bibfnamefont {S.}~\bibnamefont {Im}}, \bibinfo {author}
  {\bibfnamefont {H.~K.}\ \bibnamefont {Seo}}, \bibinfo {author} {\bibfnamefont
  {T.}~\bibnamefont {Eom}}, \bibinfo {author} {\bibfnamefont {H.}~\bibnamefont
  {Ju}}, \bibinfo {author} {\bibfnamefont {M.~K.}\ \bibnamefont {Yang}},
  \bibinfo {author} {\bibfnamefont {D.~S.}\ \bibnamefont {Jeong}},\ and\
  \bibinfo {author} {\bibfnamefont {G.~H.}\ \bibnamefont {Kim}},\ }\bibfield
  {title} {\bibinfo {title} {Purely self-rectifying memristor-based passive
  crossbar array for artificial neural network accelerators},\ }\href@noop {}
  {\bibfield  {journal} {\bibinfo  {journal} {Nature communications}\ }\textbf
  {\bibinfo {volume} {15}},\ \bibinfo {pages} {129} (\bibinfo {year}
  {2024})}\BibitemShut {NoStop}%
\bibitem [{\citenamefont {Chun}\ and\ \citenamefont {Chung}(2015)}]{Chun2015}%
  \BibitemOpen
  \bibfield  {author} {\bibinfo {author} {\bibfnamefont {H.}~\bibnamefont
  {Chun}}\ and\ \bibinfo {author} {\bibfnamefont {T.~D.}\ \bibnamefont
  {Chung}},\ }\bibfield  {title} {\bibinfo {title} {Iontronics},\ }\href
  {https://doi.org/10.1146/annurev-anchem-071114-040202} {\bibfield  {journal}
  {\bibinfo  {journal} {Annual Review of Analytical Chemistry}\ }\textbf
  {\bibinfo {volume} {8}},\ \bibinfo {pages} {441} (\bibinfo {year}
  {2015})}\BibitemShut {NoStop}%
\bibitem [{\citenamefont {Xu}\ \emph {et~al.}(2024)\citenamefont {Xu},
  \citenamefont {Zhang}, \citenamefont {Mei}, \citenamefont {Liu},
  \citenamefont {Wang},\ and\ \citenamefont {Xiao}}]{Kai}%
  \BibitemOpen
  \bibfield  {author} {\bibinfo {author} {\bibfnamefont {G.}~\bibnamefont
  {Xu}}, \bibinfo {author} {\bibfnamefont {M.}~\bibnamefont {Zhang}}, \bibinfo
  {author} {\bibfnamefont {T.}~\bibnamefont {Mei}}, \bibinfo {author}
  {\bibfnamefont {W.}~\bibnamefont {Liu}}, \bibinfo {author} {\bibfnamefont
  {L.}~\bibnamefont {Wang}},\ and\ \bibinfo {author} {\bibfnamefont
  {K.}~\bibnamefont {Xiao}},\ }\bibfield  {title} {\bibinfo {title}
  {Nanofluidic ionic memristors},\ }\href
  {https://doi.org/10.1021/acsnano.4c06467} {\bibfield  {journal} {\bibinfo
  {journal} {ACS Nano}\ }\textbf {\bibinfo {volume} {18}},\ \bibinfo {pages}
  {19423} (\bibinfo {year} {2024})},\ \bibinfo {note} {pMID: 39022809},\
  \Eprint {https://arxiv.org/abs/https://doi.org/10.1021/acsnano.4c06467}
  {https://doi.org/10.1021/acsnano.4c06467} \BibitemShut {NoStop}%
\bibitem [{\citenamefont {Feng}\ \emph {et~al.}(2010)\citenamefont {Feng},
  \citenamefont {Liu}, \citenamefont {Wu},\ and\ \citenamefont
  {Wang}}]{feng2010impedance}%
  \BibitemOpen
  \bibfield  {author} {\bibinfo {author} {\bibfnamefont {J.}~\bibnamefont
  {Feng}}, \bibinfo {author} {\bibfnamefont {J.}~\bibnamefont {Liu}}, \bibinfo
  {author} {\bibfnamefont {B.}~\bibnamefont {Wu}},\ and\ \bibinfo {author}
  {\bibfnamefont {G.}~\bibnamefont {Wang}},\ }\bibfield  {title} {\bibinfo
  {title} {Impedance characteristics of amine modified single glass
  nanopores},\ }\href@noop {} {\bibfield  {journal} {\bibinfo  {journal}
  {Analytical Chemistry}\ }\textbf {\bibinfo {volume} {82}},\ \bibinfo {pages}
  {4520} (\bibinfo {year} {2010})}\BibitemShut {NoStop}%
\bibitem [{\citenamefont {Wang}\ \emph {et~al.}(2012)\citenamefont {Wang},
  \citenamefont {Kvetny}, \citenamefont {Liu}, \citenamefont {Brown},
  \citenamefont {Li},\ and\ \citenamefont {Wang}}]{wang2012transmembrane}%
  \BibitemOpen
  \bibfield  {author} {\bibinfo {author} {\bibfnamefont {D.}~\bibnamefont
  {Wang}}, \bibinfo {author} {\bibfnamefont {M.}~\bibnamefont {Kvetny}},
  \bibinfo {author} {\bibfnamefont {J.}~\bibnamefont {Liu}}, \bibinfo {author}
  {\bibfnamefont {W.}~\bibnamefont {Brown}}, \bibinfo {author} {\bibfnamefont
  {Y.}~\bibnamefont {Li}},\ and\ \bibinfo {author} {\bibfnamefont
  {G.}~\bibnamefont {Wang}},\ }\bibfield  {title} {\bibinfo {title}
  {Transmembrane potential across single conical nanopores and resulting
  memristive and memcapacitive ion transport},\ }\href@noop {} {\bibfield
  {journal} {\bibinfo  {journal} {Journal of the American Chemical Society}\
  }\textbf {\bibinfo {volume} {134}},\ \bibinfo {pages} {3651} (\bibinfo {year}
  {2012})}\BibitemShut {NoStop}%
\bibitem [{\citenamefont {Kamsma}\ \emph {et~al.}(2024)\citenamefont {Kamsma},
  \citenamefont {Kim}, \citenamefont {Kim}, \citenamefont {Boon}, \citenamefont
  {Spitoni}, \citenamefont {Park},\ and\ \citenamefont {van
  Roij}}]{kamsma2024brain}%
  \BibitemOpen
  \bibfield  {author} {\bibinfo {author} {\bibfnamefont {T.~M.}\ \bibnamefont
  {Kamsma}}, \bibinfo {author} {\bibfnamefont {J.}~\bibnamefont {Kim}},
  \bibinfo {author} {\bibfnamefont {K.}~\bibnamefont {Kim}}, \bibinfo {author}
  {\bibfnamefont {W.~Q.}\ \bibnamefont {Boon}}, \bibinfo {author}
  {\bibfnamefont {C.}~\bibnamefont {Spitoni}}, \bibinfo {author} {\bibfnamefont
  {J.}~\bibnamefont {Park}},\ and\ \bibinfo {author} {\bibfnamefont
  {R.}~\bibnamefont {van Roij}},\ }\bibfield  {title} {\bibinfo {title}
  {Brain-inspired computing with fluidic iontronic nanochannels},\ }\href@noop
  {} {\bibfield  {journal} {\bibinfo  {journal} {Proceedings of the National
  Academy of Sciences}\ }\textbf {\bibinfo {volume} {121}},\ \bibinfo {pages}
  {e2320242121} (\bibinfo {year} {2024})}\BibitemShut {NoStop}%
\bibitem [{\citenamefont {Emmerich}\ \emph {et~al.}(2024)\citenamefont
  {Emmerich}, \citenamefont {Teng}, \citenamefont {Ronceray}, \citenamefont
  {Lopriore}, \citenamefont {Chiesa}, \citenamefont {Chernev}, \citenamefont
  {Artemov}, \citenamefont {Di~Ventra}, \citenamefont {Kis},\ and\
  \citenamefont {Radenovic}}]{emmerich2024nanofluidic}%
  \BibitemOpen
  \bibfield  {author} {\bibinfo {author} {\bibfnamefont {T.}~\bibnamefont
  {Emmerich}}, \bibinfo {author} {\bibfnamefont {Y.}~\bibnamefont {Teng}},
  \bibinfo {author} {\bibfnamefont {N.}~\bibnamefont {Ronceray}}, \bibinfo
  {author} {\bibfnamefont {E.}~\bibnamefont {Lopriore}}, \bibinfo {author}
  {\bibfnamefont {R.}~\bibnamefont {Chiesa}}, \bibinfo {author} {\bibfnamefont
  {A.}~\bibnamefont {Chernev}}, \bibinfo {author} {\bibfnamefont
  {V.}~\bibnamefont {Artemov}}, \bibinfo {author} {\bibfnamefont
  {M.}~\bibnamefont {Di~Ventra}}, \bibinfo {author} {\bibfnamefont
  {A.}~\bibnamefont {Kis}},\ and\ \bibinfo {author} {\bibfnamefont
  {A.}~\bibnamefont {Radenovic}},\ }\bibfield  {title} {\bibinfo {title}
  {Nanofluidic logic with mechano--ionic memristive switches},\ }\href@noop {}
  {\bibfield  {journal} {\bibinfo  {journal} {Nature Electronics}\ }\textbf
  {\bibinfo {volume} {7}},\ \bibinfo {pages} {271} (\bibinfo {year}
  {2024})}\BibitemShut {NoStop}%
\bibitem [{\citenamefont {Wang}\ \emph {et~al.}(2024)\citenamefont {Wang},
  \citenamefont {Liang}, \citenamefont {Ma}, \citenamefont {Shi},\ and\
  \citenamefont {Xie}}]{Wang2024}%
  \BibitemOpen
  \bibfield  {author} {\bibinfo {author} {\bibfnamefont {W.}~\bibnamefont
  {Wang}}, \bibinfo {author} {\bibfnamefont {Y.}~\bibnamefont {Liang}},
  \bibinfo {author} {\bibfnamefont {Y.}~\bibnamefont {Ma}}, \bibinfo {author}
  {\bibfnamefont {D.}~\bibnamefont {Shi}},\ and\ \bibinfo {author}
  {\bibfnamefont {Y.}~\bibnamefont {Xie}},\ }\bibfield  {title} {\bibinfo
  {title} {Memristive characteristics in an asymmetrically charged
  nanochannel},\ }\href {https://doi.org/10.1021/acs.jpclett.4c00488}
  {\bibfield  {journal} {\bibinfo  {journal} {The Journal of Physical Chemistry
  Letters}\ }\textbf {\bibinfo {volume} {15}},\ \bibinfo {pages} {6852}
  (\bibinfo {year} {2024})}\BibitemShut {NoStop}%
\bibitem [{\citenamefont {Leong}\ \emph {et~al.}(2020)\citenamefont {Leong},
  \citenamefont {Tsutsui}, \citenamefont {Murayama}, \citenamefont {Hayashida},
  \citenamefont {He},\ and\ \citenamefont {Taniguchi}}]{leong2020quasi}%
  \BibitemOpen
  \bibfield  {author} {\bibinfo {author} {\bibfnamefont {I.~W.}\ \bibnamefont
  {Leong}}, \bibinfo {author} {\bibfnamefont {M.}~\bibnamefont {Tsutsui}},
  \bibinfo {author} {\bibfnamefont {S.}~\bibnamefont {Murayama}}, \bibinfo
  {author} {\bibfnamefont {T.}~\bibnamefont {Hayashida}}, \bibinfo {author}
  {\bibfnamefont {Y.}~\bibnamefont {He}},\ and\ \bibinfo {author}
  {\bibfnamefont {M.}~\bibnamefont {Taniguchi}},\ }\bibfield  {title} {\bibinfo
  {title} {Quasi-stable salt gradient and resistive switching in solid-state
  nanopores},\ }\href@noop {} {\bibfield  {journal} {\bibinfo  {journal} {ACS
  Applied Materials \& Interfaces}\ }\textbf {\bibinfo {volume} {12}},\
  \bibinfo {pages} {52175} (\bibinfo {year} {2020})}\BibitemShut {NoStop}%
\bibitem [{\citenamefont {Smirnov}\ \emph {et~al.}(2011)\citenamefont
  {Smirnov}, \citenamefont {Vlassiouk},\ and\ \citenamefont
  {Lavrik}}]{smirnov2011voltage}%
  \BibitemOpen
  \bibfield  {author} {\bibinfo {author} {\bibfnamefont {S.~N.}\ \bibnamefont
  {Smirnov}}, \bibinfo {author} {\bibfnamefont {I.~V.}\ \bibnamefont
  {Vlassiouk}},\ and\ \bibinfo {author} {\bibfnamefont {N.~V.}\ \bibnamefont
  {Lavrik}},\ }\bibfield  {title} {\bibinfo {title} {Voltage-gated hydrophobic
  nanopores},\ }\href@noop {} {\bibfield  {journal} {\bibinfo  {journal} {Acs
  Nano}\ }\textbf {\bibinfo {volume} {5}},\ \bibinfo {pages} {7453} (\bibinfo
  {year} {2011})}\BibitemShut {NoStop}%
\bibitem [{\citenamefont {Powell}\ \emph {et~al.}(2011)\citenamefont {Powell},
  \citenamefont {Cleary}, \citenamefont {Davenport}, \citenamefont {Shea},\
  and\ \citenamefont {Siwy}}]{powell2011electric}%
  \BibitemOpen
  \bibfield  {author} {\bibinfo {author} {\bibfnamefont {M.~R.}\ \bibnamefont
  {Powell}}, \bibinfo {author} {\bibfnamefont {L.}~\bibnamefont {Cleary}},
  \bibinfo {author} {\bibfnamefont {M.}~\bibnamefont {Davenport}}, \bibinfo
  {author} {\bibfnamefont {K.~J.}\ \bibnamefont {Shea}},\ and\ \bibinfo
  {author} {\bibfnamefont {Z.~S.}\ \bibnamefont {Siwy}},\ }\bibfield  {title}
  {\bibinfo {title} {Electric-field-induced wetting and dewetting in single
  hydrophobic nanopores},\ }\href@noop {} {\bibfield  {journal} {\bibinfo
  {journal} {Nature nanotechnology}\ }\textbf {\bibinfo {volume} {6}},\
  \bibinfo {pages} {798} (\bibinfo {year} {2011})}\BibitemShut {NoStop}%
\bibitem [{\citenamefont {Tuszynski}\ \emph {et~al.}(2020)\citenamefont
  {Tuszynski}, \citenamefont {Friesen}, \citenamefont {Freedman}, \citenamefont
  {Sbitnev}, \citenamefont {Kim}, \citenamefont {Santelices}, \citenamefont
  {Kalra}, \citenamefont {Patel}, \citenamefont {Shankar},\ and\ \citenamefont
  {Chua}}]{tuszynski2020microtubules}%
  \BibitemOpen
  \bibfield  {author} {\bibinfo {author} {\bibfnamefont {J.~A.}\ \bibnamefont
  {Tuszynski}}, \bibinfo {author} {\bibfnamefont {D.}~\bibnamefont {Friesen}},
  \bibinfo {author} {\bibfnamefont {H.}~\bibnamefont {Freedman}}, \bibinfo
  {author} {\bibfnamefont {V.~I.}\ \bibnamefont {Sbitnev}}, \bibinfo {author}
  {\bibfnamefont {H.}~\bibnamefont {Kim}}, \bibinfo {author} {\bibfnamefont
  {I.}~\bibnamefont {Santelices}}, \bibinfo {author} {\bibfnamefont {A.~P.}\
  \bibnamefont {Kalra}}, \bibinfo {author} {\bibfnamefont {S.~D.}\ \bibnamefont
  {Patel}}, \bibinfo {author} {\bibfnamefont {K.}~\bibnamefont {Shankar}},\
  and\ \bibinfo {author} {\bibfnamefont {L.~O.}\ \bibnamefont {Chua}},\
  }\bibfield  {title} {\bibinfo {title} {Microtubules as sub-cellular
  memristors},\ }\href@noop {} {\bibfield  {journal} {\bibinfo  {journal}
  {Scientific reports}\ }\textbf {\bibinfo {volume} {10}},\ \bibinfo {pages}
  {2108} (\bibinfo {year} {2020})}\BibitemShut {NoStop}%
\bibitem [{\citenamefont {Zhou}\ \emph {et~al.}(2024)\citenamefont {Zhou},
  \citenamefont {Zong}, \citenamefont {Wang}, \citenamefont {Sun},
  \citenamefont {Shi}, \citenamefont {Wang}, \citenamefont {Du},\ and\
  \citenamefont {Xie}}]{zhou2024nanofluidic}%
  \BibitemOpen
  \bibfield  {author} {\bibinfo {author} {\bibfnamefont {X.}~\bibnamefont
  {Zhou}}, \bibinfo {author} {\bibfnamefont {Y.}~\bibnamefont {Zong}}, \bibinfo
  {author} {\bibfnamefont {Y.}~\bibnamefont {Wang}}, \bibinfo {author}
  {\bibfnamefont {M.}~\bibnamefont {Sun}}, \bibinfo {author} {\bibfnamefont
  {D.}~\bibnamefont {Shi}}, \bibinfo {author} {\bibfnamefont {W.}~\bibnamefont
  {Wang}}, \bibinfo {author} {\bibfnamefont {G.}~\bibnamefont {Du}},\ and\
  \bibinfo {author} {\bibfnamefont {Y.}~\bibnamefont {Xie}},\ }\bibfield
  {title} {\bibinfo {title} {Nanofluidic memristor based on the elastic
  deformation of nanopores with nanoparticle adsorption},\ }\href@noop {}
  {\bibfield  {journal} {\bibinfo  {journal} {National Science Review}\
  }\textbf {\bibinfo {volume} {11}},\ \bibinfo {pages} {nwad216} (\bibinfo
  {year} {2024})}\BibitemShut {NoStop}%
\bibitem [{\citenamefont {Robin}\ \emph {et~al.}(2023)\citenamefont {Robin},
  \citenamefont {Emmerich}, \citenamefont {Ismail}, \citenamefont {Nigu{\`e}s},
  \citenamefont {You}, \citenamefont {Nam}, \citenamefont {Keerthi},
  \citenamefont {Siria}, \citenamefont {Geim}, \citenamefont {Radha} \emph
  {et~al.}}]{robin2023long}%
  \BibitemOpen
  \bibfield  {author} {\bibinfo {author} {\bibfnamefont {P.}~\bibnamefont
  {Robin}}, \bibinfo {author} {\bibfnamefont {T.}~\bibnamefont {Emmerich}},
  \bibinfo {author} {\bibfnamefont {A.}~\bibnamefont {Ismail}}, \bibinfo
  {author} {\bibfnamefont {A.}~\bibnamefont {Nigu{\`e}s}}, \bibinfo {author}
  {\bibfnamefont {Y.}~\bibnamefont {You}}, \bibinfo {author} {\bibfnamefont
  {G.-H.}\ \bibnamefont {Nam}}, \bibinfo {author} {\bibfnamefont
  {A.}~\bibnamefont {Keerthi}}, \bibinfo {author} {\bibfnamefont
  {A.}~\bibnamefont {Siria}}, \bibinfo {author} {\bibfnamefont
  {A.}~\bibnamefont {Geim}}, \bibinfo {author} {\bibfnamefont {B.}~\bibnamefont
  {Radha}}, \emph {et~al.},\ }\bibfield  {title} {\bibinfo {title} {Long-term
  memory and synapse-like dynamics in two-dimensional nanofluidic channels},\
  }\href@noop {} {\bibfield  {journal} {\bibinfo  {journal} {Science}\ }\textbf
  {\bibinfo {volume} {379}},\ \bibinfo {pages} {161} (\bibinfo {year}
  {2023})}\BibitemShut {NoStop}%
\bibitem [{\citenamefont {Bose}\ \emph {et~al.}(2015)\citenamefont {Bose},
  \citenamefont {Lawrence}, \citenamefont {Liu}, \citenamefont {Makarenko},
  \citenamefont {van Damme}, \citenamefont {Broersma},\ and\ \citenamefont
  {van~der Wiel}}]{Bose2015}%
  \BibitemOpen
  \bibfield  {author} {\bibinfo {author} {\bibfnamefont {S.~K.}\ \bibnamefont
  {Bose}}, \bibinfo {author} {\bibfnamefont {C.~P.}\ \bibnamefont {Lawrence}},
  \bibinfo {author} {\bibfnamefont {Z.}~\bibnamefont {Liu}}, \bibinfo {author}
  {\bibfnamefont {K.~S.}\ \bibnamefont {Makarenko}}, \bibinfo {author}
  {\bibfnamefont {R.~M.~J.}\ \bibnamefont {van Damme}}, \bibinfo {author}
  {\bibfnamefont {H.~J.}\ \bibnamefont {Broersma}},\ and\ \bibinfo {author}
  {\bibfnamefont {W.~G.}\ \bibnamefont {van~der Wiel}},\ }\bibfield  {title}
  {\bibinfo {title} {Evolution of a designless nanoparticle network into
  reconfigurable boolean logic},\ }\href
  {https://doi.org/10.1038/nnano.2015.207} {\bibfield  {journal} {\bibinfo
  {journal} {Nature Nanotechnology}\ }\textbf {\bibinfo {volume} {10}},\
  \bibinfo {pages} {1048} (\bibinfo {year} {2015})}\BibitemShut {NoStop}%
\bibitem [{\citenamefont {Sheng}\ \emph {et~al.}(2017)\citenamefont {Sheng},
  \citenamefont {Xie}, \citenamefont {Li}, \citenamefont {Wang},\ and\
  \citenamefont {Xue}}]{sheng2017transporting}%
  \BibitemOpen
  \bibfield  {author} {\bibinfo {author} {\bibfnamefont {Q.}~\bibnamefont
  {Sheng}}, \bibinfo {author} {\bibfnamefont {Y.}~\bibnamefont {Xie}}, \bibinfo
  {author} {\bibfnamefont {J.}~\bibnamefont {Li}}, \bibinfo {author}
  {\bibfnamefont {X.}~\bibnamefont {Wang}},\ and\ \bibinfo {author}
  {\bibfnamefont {J.}~\bibnamefont {Xue}},\ }\bibfield  {title} {\bibinfo
  {title} {Transporting an ionic-liquid/water mixture in a conical nanochannel:
  a nanofluidic memristor},\ }\href@noop {} {\bibfield  {journal} {\bibinfo
  {journal} {Chemical Communications}\ }\textbf {\bibinfo {volume} {53}},\
  \bibinfo {pages} {6125} (\bibinfo {year} {2017})}\BibitemShut {NoStop}%
\bibitem [{\citenamefont {Zhang}\ \emph {et~al.}(2019)\citenamefont {Zhang},
  \citenamefont {Xia}, \citenamefont {Zhuge}, \citenamefont {Zhou},
  \citenamefont {Wang}, \citenamefont {Dong}, \citenamefont {Fu}, \citenamefont
  {Yang}, \citenamefont {Li}, \citenamefont {He} \emph
  {et~al.}}]{zhang2019nanochannel}%
  \BibitemOpen
  \bibfield  {author} {\bibinfo {author} {\bibfnamefont {P.}~\bibnamefont
  {Zhang}}, \bibinfo {author} {\bibfnamefont {M.}~\bibnamefont {Xia}}, \bibinfo
  {author} {\bibfnamefont {F.}~\bibnamefont {Zhuge}}, \bibinfo {author}
  {\bibfnamefont {Y.}~\bibnamefont {Zhou}}, \bibinfo {author} {\bibfnamefont
  {Z.}~\bibnamefont {Wang}}, \bibinfo {author} {\bibfnamefont {B.}~\bibnamefont
  {Dong}}, \bibinfo {author} {\bibfnamefont {Y.}~\bibnamefont {Fu}}, \bibinfo
  {author} {\bibfnamefont {K.}~\bibnamefont {Yang}}, \bibinfo {author}
  {\bibfnamefont {Y.}~\bibnamefont {Li}}, \bibinfo {author} {\bibfnamefont
  {Y.}~\bibnamefont {He}}, \emph {et~al.},\ }\bibfield  {title} {\bibinfo
  {title} {Nanochannel-based transport in an interfacial memristor can emulate
  the analog weight modulation of synapses},\ }\href@noop {} {\bibfield
  {journal} {\bibinfo  {journal} {Nano Letters}\ }\textbf {\bibinfo {volume}
  {19}},\ \bibinfo {pages} {4279} (\bibinfo {year} {2019})}\BibitemShut
  {NoStop}%
\bibitem [{\citenamefont {Xiong}\ \emph {et~al.}(2023)\citenamefont {Xiong},
  \citenamefont {Li}, \citenamefont {He}, \citenamefont {Xie}, \citenamefont
  {Zong}, \citenamefont {Jiang}, \citenamefont {Ma}, \citenamefont {Wu},
  \citenamefont {Fei}, \citenamefont {Yu} \emph
  {et~al.}}]{xiong2023neuromorphic}%
  \BibitemOpen
  \bibfield  {author} {\bibinfo {author} {\bibfnamefont {T.}~\bibnamefont
  {Xiong}}, \bibinfo {author} {\bibfnamefont {C.}~\bibnamefont {Li}}, \bibinfo
  {author} {\bibfnamefont {X.}~\bibnamefont {He}}, \bibinfo {author}
  {\bibfnamefont {B.}~\bibnamefont {Xie}}, \bibinfo {author} {\bibfnamefont
  {J.}~\bibnamefont {Zong}}, \bibinfo {author} {\bibfnamefont {Y.}~\bibnamefont
  {Jiang}}, \bibinfo {author} {\bibfnamefont {W.}~\bibnamefont {Ma}}, \bibinfo
  {author} {\bibfnamefont {F.}~\bibnamefont {Wu}}, \bibinfo {author}
  {\bibfnamefont {J.}~\bibnamefont {Fei}}, \bibinfo {author} {\bibfnamefont
  {P.}~\bibnamefont {Yu}}, \emph {et~al.},\ }\bibfield  {title} {\bibinfo
  {title} {Neuromorphic functions with a polyelectrolyte-confined fluidic
  memristor},\ }\href@noop {} {\bibfield  {journal} {\bibinfo  {journal}
  {Science}\ }\textbf {\bibinfo {volume} {379}},\ \bibinfo {pages} {156}
  (\bibinfo {year} {2023})}\BibitemShut {NoStop}%
\bibitem [{\citenamefont {van Doremaele}\ \emph {et~al.}(2024)\citenamefont
  {van Doremaele}, \citenamefont {Stevens}, \citenamefont {Ringeling},
  \citenamefont {Spolaor}, \citenamefont {Fattori},\ and\ \citenamefont {van~de
  Burgt}}]{Doremaele2024}%
  \BibitemOpen
  \bibfield  {author} {\bibinfo {author} {\bibfnamefont {E.~R.~W.}\
  \bibnamefont {van Doremaele}}, \bibinfo {author} {\bibfnamefont
  {T.}~\bibnamefont {Stevens}}, \bibinfo {author} {\bibfnamefont
  {S.}~\bibnamefont {Ringeling}}, \bibinfo {author} {\bibfnamefont
  {S.}~\bibnamefont {Spolaor}}, \bibinfo {author} {\bibfnamefont
  {M.}~\bibnamefont {Fattori}},\ and\ \bibinfo {author} {\bibfnamefont
  {Y.}~\bibnamefont {van~de Burgt}},\ }\bibfield  {title} {\bibinfo {title}
  {Hardware implementation of backpropagation using progressive gradient
  descent for in situ training of multilayer neural networks},\ }\bibfield
  {journal} {\bibinfo  {journal} {Science Advances}\ }\textbf {\bibinfo
  {volume} {10}},\ \href {https://doi.org/10.1126/sciadv.ado8999}
  {10.1126/sciadv.ado8999} (\bibinfo {year} {2024})\BibitemShut {NoStop}%
\bibitem [{\citenamefont {Sabbagh}\ \emph {et~al.}(2023)\citenamefont
  {Sabbagh}, \citenamefont {Fraiman}, \citenamefont {Fish},\ and\ \citenamefont
  {Yossifon}}]{sabbagh2023designing}%
  \BibitemOpen
  \bibfield  {author} {\bibinfo {author} {\bibfnamefont {B.}~\bibnamefont
  {Sabbagh}}, \bibinfo {author} {\bibfnamefont {N.~E.}\ \bibnamefont
  {Fraiman}}, \bibinfo {author} {\bibfnamefont {A.}~\bibnamefont {Fish}},\ and\
  \bibinfo {author} {\bibfnamefont {G.}~\bibnamefont {Yossifon}},\ }\bibfield
  {title} {\bibinfo {title} {Designing with iontronic logic gates -- from a
  single polyelectrolyte diode to an integrated ionic circuit},\ }\href@noop {}
  {\bibfield  {journal} {\bibinfo  {journal} {ACS Applied Materials \&
  Interfaces}\ }\textbf {\bibinfo {volume} {15}},\ \bibinfo {pages} {23361}
  (\bibinfo {year} {2023})}\BibitemShut {NoStop}%
\bibitem [{\citenamefont {Li}\ \emph {et~al.}(2023)\citenamefont {Li},
  \citenamefont {Li}, \citenamefont {Zhang}, \citenamefont {Hu},\ and\
  \citenamefont {Li}}]{Li2023}%
  \BibitemOpen
  \bibfield  {author} {\bibinfo {author} {\bibfnamefont {J.}~\bibnamefont
  {Li}}, \bibinfo {author} {\bibfnamefont {M.}~\bibnamefont {Li}}, \bibinfo
  {author} {\bibfnamefont {K.}~\bibnamefont {Zhang}}, \bibinfo {author}
  {\bibfnamefont {L.}~\bibnamefont {Hu}},\ and\ \bibinfo {author}
  {\bibfnamefont {D.}~\bibnamefont {Li}},\ }\bibfield  {title} {\bibinfo
  {title} {High‐performance integrated iontronic circuits based on single
  nano/microchannels},\ }\bibfield  {journal} {\bibinfo  {journal} {Small}\
  }\textbf {\bibinfo {volume} {19}},\ \href
  {https://doi.org/10.1002/smll.202208079} {10.1002/smll.202208079} (\bibinfo
  {year} {2023})\BibitemShut {NoStop}%
\bibitem [{\citenamefont {Zhang}\ \emph {et~al.}(2024)\citenamefont {Zhang},
  \citenamefont {Tan}, \citenamefont {Toepfer}, \citenamefont {Lu},\ and\
  \citenamefont {Bayley}}]{zhang2024microscale}%
  \BibitemOpen
  \bibfield  {author} {\bibinfo {author} {\bibfnamefont {Y.}~\bibnamefont
  {Zhang}}, \bibinfo {author} {\bibfnamefont {C.~M.}\ \bibnamefont {Tan}},
  \bibinfo {author} {\bibfnamefont {C.~N.}\ \bibnamefont {Toepfer}}, \bibinfo
  {author} {\bibfnamefont {X.}~\bibnamefont {Lu}},\ and\ \bibinfo {author}
  {\bibfnamefont {H.}~\bibnamefont {Bayley}},\ }\bibfield  {title} {\bibinfo
  {title} {Microscale droplet assembly enables biocompatible multifunctional
  modular iontronics},\ }\href@noop {} {\bibfield  {journal} {\bibinfo
  {journal} {Science}\ }\textbf {\bibinfo {volume} {386}},\ \bibinfo {pages}
  {1024} (\bibinfo {year} {2024})}\BibitemShut {NoStop}%
\bibitem [{\citenamefont {Portillo}\ \emph {et~al.}(2024)\citenamefont
  {Portillo}, \citenamefont {Cervera}, \citenamefont {Mafe},\ and\
  \citenamefont {Ramirez}}]{Portillo2024}%
  \BibitemOpen
  \bibfield  {author} {\bibinfo {author} {\bibfnamefont {S.}~\bibnamefont
  {Portillo}}, \bibinfo {author} {\bibfnamefont {J.}~\bibnamefont {Cervera}},
  \bibinfo {author} {\bibfnamefont {S.}~\bibnamefont {Mafe}},\ and\ \bibinfo
  {author} {\bibfnamefont {P.}~\bibnamefont {Ramirez}},\ }\bibfield  {title}
  {\bibinfo {title} {Reversible logic with a nanofluidic memristor},\ }\href
  {https://doi.org/10.1103/physreve.110.065101} {\bibfield  {journal} {\bibinfo
   {journal} {Physical Review E}\ }\textbf {\bibinfo {volume} {110}},\ \bibinfo
  {pages} {065101} (\bibinfo {year} {2024})}\BibitemShut {NoStop}%
\bibitem [{\citenamefont {Ling}\ \emph {et~al.}(2024)\citenamefont {Ling},
  \citenamefont {Yu}, \citenamefont {Guo}, \citenamefont {Bian}, \citenamefont
  {Wang}, \citenamefont {Wang}, \citenamefont {Hou},\ and\ \citenamefont
  {Hou}}]{Ling2024}%
  \BibitemOpen
  \bibfield  {author} {\bibinfo {author} {\bibfnamefont {Y.}~\bibnamefont
  {Ling}}, \bibinfo {author} {\bibfnamefont {L.}~\bibnamefont {Yu}}, \bibinfo
  {author} {\bibfnamefont {Z.}~\bibnamefont {Guo}}, \bibinfo {author}
  {\bibfnamefont {F.}~\bibnamefont {Bian}}, \bibinfo {author} {\bibfnamefont
  {Y.}~\bibnamefont {Wang}}, \bibinfo {author} {\bibfnamefont {X.}~\bibnamefont
  {Wang}}, \bibinfo {author} {\bibfnamefont {Y.}~\bibnamefont {Hou}},\ and\
  \bibinfo {author} {\bibfnamefont {X.}~\bibnamefont {Hou}},\ }\bibfield
  {title} {\bibinfo {title} {Single-pore nanofluidic logic memristor with
  reconfigurable synaptic functions and designable combinations},\ }\href
  {https://doi.org/10.1021/jacs.4c01218} {\bibfield  {journal} {\bibinfo
  {journal} {Journal of the American Chemical Society}\ }\textbf {\bibinfo
  {volume} {146}},\ \bibinfo {pages} {14558} (\bibinfo {year}
  {2024})}\BibitemShut {NoStop}%
\bibitem [{\citenamefont {Liu}\ \emph {et~al.}(2024)\citenamefont {Liu},
  \citenamefont {Mei}, \citenamefont {Cao}, \citenamefont {Li}, \citenamefont
  {Wu}, \citenamefont {Wang}, \citenamefont {Xu}, \citenamefont {Chen},
  \citenamefont {Zhou}, \citenamefont {Wang}, \citenamefont {Xue},
  \citenamefont {Yu}, \citenamefont {Kong}, \citenamefont {Chen}, \citenamefont
  {Tu},\ and\ \citenamefont {Xiao}}]{Liu2024}%
  \BibitemOpen
  \bibfield  {author} {\bibinfo {author} {\bibfnamefont {W.}~\bibnamefont
  {Liu}}, \bibinfo {author} {\bibfnamefont {T.}~\bibnamefont {Mei}}, \bibinfo
  {author} {\bibfnamefont {Z.}~\bibnamefont {Cao}}, \bibinfo {author}
  {\bibfnamefont {C.}~\bibnamefont {Li}}, \bibinfo {author} {\bibfnamefont
  {Y.}~\bibnamefont {Wu}}, \bibinfo {author} {\bibfnamefont {L.}~\bibnamefont
  {Wang}}, \bibinfo {author} {\bibfnamefont {G.}~\bibnamefont {Xu}}, \bibinfo
  {author} {\bibfnamefont {Y.}~\bibnamefont {Chen}}, \bibinfo {author}
  {\bibfnamefont {Y.}~\bibnamefont {Zhou}}, \bibinfo {author} {\bibfnamefont
  {S.}~\bibnamefont {Wang}}, \bibinfo {author} {\bibfnamefont {Y.}~\bibnamefont
  {Xue}}, \bibinfo {author} {\bibfnamefont {Y.}~\bibnamefont {Yu}}, \bibinfo
  {author} {\bibfnamefont {X.-Y.}\ \bibnamefont {Kong}}, \bibinfo {author}
  {\bibfnamefont {R.}~\bibnamefont {Chen}}, \bibinfo {author} {\bibfnamefont
  {B.}~\bibnamefont {Tu}},\ and\ \bibinfo {author} {\bibfnamefont
  {K.}~\bibnamefont {Xiao}},\ }\bibfield  {title} {\bibinfo {title}
  {Bioinspired carbon nanotube–based nanofluidic ionic transistor with
  ultrahigh switching capabilities for logic circuits},\ }\bibfield  {journal}
  {\bibinfo  {journal} {Science Advances}\ }\textbf {\bibinfo {volume} {10}},\
  \href {https://doi.org/10.1126/sciadv.adj7867} {10.1126/sciadv.adj7867}
  (\bibinfo {year} {2024})\BibitemShut {NoStop}%
\bibitem [{\citenamefont {Kamsma}\ \emph
  {et~al.}(2023{\natexlab{a}})\citenamefont {Kamsma}, \citenamefont {Boon},
  \citenamefont {ter Rele}, \citenamefont {Spitoni},\ and\ \citenamefont {van
  Roij}}]{Kamsma2023}%
  \BibitemOpen
  \bibfield  {author} {\bibinfo {author} {\bibfnamefont {T.}~\bibnamefont
  {Kamsma}}, \bibinfo {author} {\bibfnamefont {W.}~\bibnamefont {Boon}},
  \bibinfo {author} {\bibfnamefont {T.}~\bibnamefont {ter Rele}}, \bibinfo
  {author} {\bibfnamefont {C.}~\bibnamefont {Spitoni}},\ and\ \bibinfo {author}
  {\bibfnamefont {R.}~\bibnamefont {van Roij}},\ }\bibfield  {title} {\bibinfo
  {title} {Iontronic neuromorphic signaling with conical microfluidic
  memristors},\ }\href {https://doi.org/10.1103/physrevlett.130.268401}
  {\bibfield  {journal} {\bibinfo  {journal} {Physical Review Letters}\
  }\textbf {\bibinfo {volume} {130}},\ \bibinfo {pages} {268401} (\bibinfo
  {year} {2023}{\natexlab{a}})}\BibitemShut {NoStop}%
\bibitem [{\citenamefont {Shinriki}\ \emph {et~al.}(1981)\citenamefont
  {Shinriki}, \citenamefont {Yamamoto},\ and\ \citenamefont
  {Mori}}]{Shinriki1981}%
  \BibitemOpen
  \bibfield  {author} {\bibinfo {author} {\bibfnamefont {M.}~\bibnamefont
  {Shinriki}}, \bibinfo {author} {\bibfnamefont {M.}~\bibnamefont {Yamamoto}},\
  and\ \bibinfo {author} {\bibfnamefont {S.}~\bibnamefont {Mori}},\ }\bibfield
  {title} {\bibinfo {title} {Multimode oscillations in a modified van der pol
  oscillator containing a positive nonlinear conductance},\ }\href
  {https://doi.org/10.1109/proc.1981.11973} {\bibfield  {journal} {\bibinfo
  {journal} {Proceedings of the IEEE}\ }\textbf {\bibinfo {volume} {69}},\
  \bibinfo {pages} {394} (\bibinfo {year} {1981})}\BibitemShut {NoStop}%
\bibitem [{\citenamefont {Kitzbichler}\ \emph {et~al.}(2009)\citenamefont
  {Kitzbichler}, \citenamefont {Smith}, \citenamefont {Christensen},\ and\
  \citenamefont {Bullmore}}]{Kitzbichler2009}%
  \BibitemOpen
  \bibfield  {author} {\bibinfo {author} {\bibfnamefont {M.~G.}\ \bibnamefont
  {Kitzbichler}}, \bibinfo {author} {\bibfnamefont {M.~L.}\ \bibnamefont
  {Smith}}, \bibinfo {author} {\bibfnamefont {S.~R.}\ \bibnamefont
  {Christensen}},\ and\ \bibinfo {author} {\bibfnamefont {E.}~\bibnamefont
  {Bullmore}},\ }\bibfield  {title} {\bibinfo {title} {Broadband criticality of
  human brain network synchronization},\ }\href
  {https://doi.org/10.1371/journal.pcbi.1000314} {\bibfield  {journal}
  {\bibinfo  {journal} {PLoS Computational Biology}\ }\textbf {\bibinfo
  {volume} {5}},\ \bibinfo {pages} {e1000314} (\bibinfo {year}
  {2009})}\BibitemShut {NoStop}%
\bibitem [{\citenamefont {Chua}(2013)}]{chua2013memristor}%
  \BibitemOpen
  \bibfield  {author} {\bibinfo {author} {\bibfnamefont {L.}~\bibnamefont
  {Chua}},\ }\bibfield  {title} {\bibinfo {title} {Memristor, hodgkin--huxley,
  and edge of chaos},\ }\href@noop {} {\bibfield  {journal} {\bibinfo
  {journal} {Nanotechnology}\ }\textbf {\bibinfo {volume} {24}},\ \bibinfo
  {pages} {383001} (\bibinfo {year} {2013})}\BibitemShut {NoStop}%
\bibitem [{\citenamefont {Jubin}\ \emph {et~al.}(2018)\citenamefont {Jubin},
  \citenamefont {Poggioli}, \citenamefont {Siria},\ and\ \citenamefont
  {Bocquet}}]{jubin2018dramatic}%
  \BibitemOpen
  \bibfield  {author} {\bibinfo {author} {\bibfnamefont {L.}~\bibnamefont
  {Jubin}}, \bibinfo {author} {\bibfnamefont {A.}~\bibnamefont {Poggioli}},
  \bibinfo {author} {\bibfnamefont {A.}~\bibnamefont {Siria}},\ and\ \bibinfo
  {author} {\bibfnamefont {L.}~\bibnamefont {Bocquet}},\ }\bibfield  {title}
  {\bibinfo {title} {Dramatic pressure-sensitive ion conduction in conical
  nanopores},\ }\href@noop {} {\bibfield  {journal} {\bibinfo  {journal}
  {Proceedings of the National Academy of Sciences}\ }\textbf {\bibinfo
  {volume} {115}},\ \bibinfo {pages} {4063} (\bibinfo {year}
  {2018})}\BibitemShut {NoStop}%
\bibitem [{\citenamefont {Boon}\ \emph {et~al.}(2022)\citenamefont {Boon},
  \citenamefont {Veenstra}, \citenamefont {Dijkstra},\ and\ \citenamefont {van
  Roij}}]{boon2022pressure}%
  \BibitemOpen
  \bibfield  {author} {\bibinfo {author} {\bibfnamefont {W.~Q.}\ \bibnamefont
  {Boon}}, \bibinfo {author} {\bibfnamefont {T.~E.}\ \bibnamefont {Veenstra}},
  \bibinfo {author} {\bibfnamefont {M.}~\bibnamefont {Dijkstra}},\ and\
  \bibinfo {author} {\bibfnamefont {R.}~\bibnamefont {van Roij}},\ }\bibfield
  {title} {\bibinfo {title} {Pressure-sensitive ion conduction in a conical
  channel: Optimal pressure and geometry},\ }\href@noop {} {\bibfield
  {journal} {\bibinfo  {journal} {Physics of Fluids}\ }\textbf {\bibinfo
  {volume} {34}} (\bibinfo {year} {2022})}\BibitemShut {NoStop}%
\bibitem [{Note1()}]{Note1}%
  \BibitemOpen
  \bibinfo {note} {This is only an approximation of the steady-state
  conductance, but Kamsma {\protect \em et al.} \ found that it is reasonable
  within the applied voltage range~\cite {Kamsma2023,Kamsma2023a}}\BibitemShut
  {NoStop}%
\bibitem [{\citenamefont {Holters}\ and\ \citenamefont {Zölzer}(2015)}]{ACME}%
  \BibitemOpen
  \bibfield  {author} {\bibinfo {author} {\bibfnamefont {M.}~\bibnamefont
  {Holters}}\ and\ \bibinfo {author} {\bibfnamefont {U.}~\bibnamefont
  {Zölzer}},\ }\bibfield  {title} {\bibinfo {title} {A generalized method for
  the derivation of non-linear state-space models from circuit schematics},\
  }in\ \href {https://doi.org/10.1109/EUSIPCO.2015.7362548} {\emph {\bibinfo
  {booktitle} {2015 23rd European Signal Processing Conference (EUSIPCO)}}}\
  (\bibinfo {year} {2015})\ pp.\ \bibinfo {pages} {1073--1077}\BibitemShut
  {NoStop}%
\bibitem [{\citenamefont {Stuhlmüller}\ \emph {et~al.}(2025)\citenamefont
  {Stuhlmüller}, \citenamefont {van Roij},\ and\ \citenamefont
  {Dijkstra}}]{code}%
  \BibitemOpen
  \bibfield  {author} {\bibinfo {author} {\bibfnamefont {N.~C.~X.}\
  \bibnamefont {Stuhlmüller}}, \bibinfo {author} {\bibfnamefont
  {R.}~\bibnamefont {van Roij}},\ and\ \bibinfo {author} {\bibfnamefont
  {M.}~\bibnamefont {Dijkstra}},\ }\href
  {https://doi.org/10.5281/zenodo.14944227} {\bibinfo {title} {Memriki-logic:
  Publication release}} (\bibinfo {year} {2025})\BibitemShut {NoStop}%
\bibitem [{\citenamefont {Stojanović}\ \emph {et~al.}(2019)\citenamefont
  {Stojanović}, \citenamefont {Paroški}, \citenamefont {Samardžić},
  \citenamefont {Radovanović},\ and\ \citenamefont
  {Krstić}}]{Stojanovic2019}%
  \BibitemOpen
  \bibfield  {author} {\bibinfo {author} {\bibfnamefont {G.}~\bibnamefont
  {Stojanović}}, \bibinfo {author} {\bibfnamefont {M.}~\bibnamefont
  {Paroški}}, \bibinfo {author} {\bibfnamefont {N.}~\bibnamefont
  {Samardžić}}, \bibinfo {author} {\bibfnamefont {M.}~\bibnamefont
  {Radovanović}},\ and\ \bibinfo {author} {\bibfnamefont {D.}~\bibnamefont
  {Krstić}},\ }\bibfield  {title} {\bibinfo {title} {Microfluidics-based four
  fundamental electronic circuit elements resistor, inductor, capacitor and
  memristor},\ }\href {https://doi.org/10.3390/electronics8090960} {\bibfield
  {journal} {\bibinfo  {journal} {Electronics}\ }\textbf {\bibinfo {volume}
  {8}},\ \bibinfo {pages} {960} (\bibinfo {year} {2019})}\BibitemShut {NoStop}%
\bibitem [{\citenamefont {Gallardo~Hevia}\ \emph {et~al.}(2022)\citenamefont
  {Gallardo~Hevia}, \citenamefont {McCann}, \citenamefont {Bell}, \citenamefont
  {Hyun}, \citenamefont {Majidi}, \citenamefont {Bertoldi},\ and\ \citenamefont
  {Wood}}]{GallardoHevia2022}%
  \BibitemOpen
  \bibfield  {author} {\bibinfo {author} {\bibfnamefont {E.}~\bibnamefont
  {Gallardo~Hevia}}, \bibinfo {author} {\bibfnamefont {C.~M.}\ \bibnamefont
  {McCann}}, \bibinfo {author} {\bibfnamefont {M.}~\bibnamefont {Bell}},
  \bibinfo {author} {\bibfnamefont {N.-s.~P.}\ \bibnamefont {Hyun}}, \bibinfo
  {author} {\bibfnamefont {C.}~\bibnamefont {Majidi}}, \bibinfo {author}
  {\bibfnamefont {K.}~\bibnamefont {Bertoldi}},\ and\ \bibinfo {author}
  {\bibfnamefont {R.~J.}\ \bibnamefont {Wood}},\ }\bibfield  {title} {\bibinfo
  {title} {High‐gain microfluidic amplifiers: The bridge between microfluidic
  controllers and fluidic soft actuators},\ }\bibfield  {journal} {\bibinfo
  {journal} {Advanced Intelligent Systems}\ }\textbf {\bibinfo {volume} {4}},\
  \href {https://doi.org/10.1002/aisy.202200122} {10.1002/aisy.202200122}
  (\bibinfo {year} {2022})\BibitemShut {NoStop}%
\bibitem [{\citenamefont {Minsky}\ and\ \citenamefont
  {Papert}(2017)}]{minsky2017perceptrons}%
  \BibitemOpen
  \bibfield  {author} {\bibinfo {author} {\bibfnamefont {M.}~\bibnamefont
  {Minsky}}\ and\ \bibinfo {author} {\bibfnamefont {S.~A.}\ \bibnamefont
  {Papert}},\ }\href@noop {} {\emph {\bibinfo {title} {Perceptrons, reissue of
  the 1988 expanded edition with a new foreword by L{\'e}on Bottou: an
  introduction to computational geometry}}}\ (\bibinfo  {publisher} {MIT
  press},\ \bibinfo {year} {2017})\BibitemShut {NoStop}%
\bibitem [{\citenamefont {Shockley}(1949)}]{Shockley1949}%
  \BibitemOpen
  \bibfield  {author} {\bibinfo {author} {\bibfnamefont {W.}~\bibnamefont
  {Shockley}},\ }\bibfield  {title} {\bibinfo {title} {The theory of p-n
  junctions in semiconductors and p-n junction transistors},\ }\href
  {https://doi.org/10.1002/j.1538-7305.1949.tb03645.x} {\bibfield  {journal}
  {\bibinfo  {journal} {Bell System Technical Journal}\ }\textbf {\bibinfo
  {volume} {28}},\ \bibinfo {pages} {435} (\bibinfo {year} {1949})}\BibitemShut
  {NoStop}%
\bibitem [{\citenamefont {Stuhlmüller}(2025)}]{data}%
  \BibitemOpen
  \bibfield  {author} {\bibinfo {author} {\bibfnamefont {N.~C.~X.}\
  \bibnamefont {Stuhlmüller}},\ }\bibfield  {title} {\bibinfo {title}
  {Iontronic memristor logic gate voltage curves},\ }\href
  {https://doi.org/10.5281/zenodo.14924500} {10.5281/zenodo.14924500} (\bibinfo
  {year} {2025})\BibitemShut {NoStop}%
\bibitem [{\citenamefont {Kamsma}\ \emph
  {et~al.}(2023{\natexlab{b}})\citenamefont {Kamsma}, \citenamefont {Boon},
  \citenamefont {Spitoni},\ and\ \citenamefont {van Roij}}]{Kamsma2023a}%
  \BibitemOpen
  \bibfield  {author} {\bibinfo {author} {\bibfnamefont {T.}~\bibnamefont
  {Kamsma}}, \bibinfo {author} {\bibfnamefont {W.~Q.}\ \bibnamefont {Boon}},
  \bibinfo {author} {\bibfnamefont {C.}~\bibnamefont {Spitoni}},\ and\ \bibinfo
  {author} {\bibfnamefont {R.}~\bibnamefont {van Roij}},\ }\bibfield  {title}
  {\bibinfo {title} {Unveiling the capabilities of bipolar conical channels in
  neuromorphic iontronics},\ }\href {https://doi.org/10.1039/d3fd00022b}
  {\bibfield  {journal} {\bibinfo  {journal} {Faraday Discussions}\ }\textbf
  {\bibinfo {volume} {246}},\ \bibinfo {pages} {125} (\bibinfo {year}
  {2023}{\natexlab{b}})}\BibitemShut {NoStop}%
\end{thebibliography}%
\appendix

\end{document}